\documentclass[twocolumn]{aastex63}
\usepackage{hyperref}
\usepackage{amsmath}
\usepackage{natbib}

\usepackage{xcolor}
\usepackage{tabularx}

\usepackage{multirow}

\begin{document}
\title{The Ages of Optically Bright Sub-Clusters in the Serpens Star-Forming Region}

\author{Xingyu Zhou}
\affiliation{Kavli Institute for Astronomy and Astrophysics, Peking University, Yiheyuan 5, Haidian Qu, 100871 Beijing, China}
\affiliation{Department of Astronomy, Peking University, Yiheyuan 5, Haidian Qu, 100871 Beijing, China}

\author{Gregory J. Herczeg}
\affiliation{Kavli Institute for Astronomy and Astrophysics, Peking University, Yiheyuan 5, Haidian Qu, 100871 Beijing, China}
\affiliation{Department of Astronomy, Peking University, Yiheyuan 5, Haidian Qu, 100871 Beijing, China}

\author{Yao Liu}
\affiliation{Purple Mountain Observatory, Chinese Academy of Sciences, 10 Yuanhua Road, Nanjing 210023, China}

\author{Min Fang}
\affiliation{Purple Mountain Observatory, Chinese Academy of Sciences, 10 Yuanhua Road, Nanjing 210023, China}

\author{Michael Kuhn}
\affiliation{California Institute of Technology, Pasadena, CA 91125, USA}

\begin{abstract}
The Serpens Molecular Cloud is one of the most active star-forming regions within 500 pc, with over one thousand of YSOs at different evolutionary stages. The ages of the member stars inform us about the star formation history of the cloud.  In this paper, we develop a spectral energy distribution (SED) fitting method for nearby evolved (diskless) young stars from members of the Pleiades to estimate their ages, with a temperature scale adopted from APOGEE spectra. When compared with literature temperatures of selected YSOs in Orion, the SED fits to cool ($<5000$ K) stars have temperatures that differ by an average of $<\sim$ 50 K and have a scatter of $\sim 210$ K for both disk-hosting and diskless stars. We then apply this method to YSOs in the Serpens Molecular Cloud to estimate ages of optical members previously identified from \textit{Gaia} DR2 astrometry data. The optical members in Serpens are concentrated in different subgroups with ages from $\sim4$ Myr to $\sim22$ Myr; the youngest clusters, W40 and Serpens South, are dusty regions that lack enough optical members to be included in this analysis.  These ages establish that the Serpens Molecular Cloud has been forming stars for much longer than has been inferred from infrared surveys.
\end{abstract}


\section{Introduction}

The story of the star-forming history of a molecular cloud is told through its dust, gas, and member stars.  The youngest members, the protostars, { are still forming} and are typically clustered in the densest regions of the cloud.  Stars formed in previous bursts are often still located in small subclusters, while some have been dispersed throughout the cloud or were born in small groups.  In many clouds the older stars are not co-located with the youngest stars, indicating that the site for star formation changes with time within a cloud complex \citep[e.g.,][]{Getman2014a,Getman2014b, Vazquez-Semadeni2018,Kounkel2018,kristensen18,Liu2021}. 

The Serpens Molecular Cloud is one of the nearest clouds with vigorous ongoing star formation, second to only Orion among active star-forming sites within 500 pc.  Historically the Serpens Main cluster \citep[see review by][]{eiroa08} and W40 (\citealt{Kuhn2010}; see also review by \citealt{rodney08}) were recognized as primary sites of star formation in the complex; mid- and far-IR and X-ray surveys revealed vigorous ongoing star formation in the deeply embedded Serpens South cluster \citep{Gutermuth2008,povich13,mallick13,konyves15,Dunham2015}.  Beyond these sub-clusters, recently-formed stars are sparse and hard to find \citep[see review by][]{prato08}.  Accurate astrometry and optical photometry from Gaia DR2 revealed the optical members, including some in new subclusters and some distributed throughout the cloud \citep{Herczeg2019}, while most candidate members that were outside the main star-forming sites have been identified as contaminants \citep[see also][]{lee21}.  The optical members that are distributed across the cloud likely formed in past epochs of star formation.

\begin{deluxetable*}{cccc}[!th]
\label{table:photometry_criteria}
\tablecaption{Photometry selection criteria}
\tablehead{\nocolhead{Survey} & \colhead{Pass-band} & \colhead{Criteria} & \colhead{Note} }
\startdata
\textit{Gaia}        & $BP,RP$     &  \multicolumn{1}{c}{\begin{tabular}[c]{@{}c@{}}\texttt{parallax\_over\_error>10 } \\                                                           \texttt{astrometric\_excess\_noise<2} \\ \texttt{visibility\_periods\_used>8} \\                                           \texttt{phot\_g\_mean\_flux\_over\_error>50}\end{tabular}}  & \\\hline
2MASS       & $J,H,K$     & \texttt{ph\_qual="AAA"}   &                                              \\ \hline
WISE       & $W1,W2$     & \multicolumn{1}{c}{\begin{tabular}[c]{@{}c@{}}\texttt{SNR>10}\\ \texttt{$\chi^2$<3}\end{tabular}}  & \begin{tabular}{p{0.8\columnwidth}} W1 and W2 are selected and treated independently. \end{tabular}    \\ \hline
Pan-STARRS  & $g,r,i,z,y$ & \multicolumn{1}{c}{\begin{tabular}[c]{@{}c@{}}\texttt{nDetections>5}\\ \texttt{gQfPerfect>0.85}\\ \texttt{rQfPerfect>0.85}\\ \texttt{iQfPerfect>0.85}\\ \texttt{zQfPerfect>0.85}\\ \texttt{imeanpsfmag - imeankronmag < 0.05} \\ \texttt{OR} \\  \texttt{rmeanpsfmag - rmeankronmag < 0.05} \end{tabular}} & \begin{tabular}{p{0.8\columnwidth}} Pan-STARRS data of a star will be used only if the data satisfy all the criteria. \end{tabular}    \\ \hline
APASS       & $B,V,g',r',i'$ & - & \begin{tabular}{p{0.8\columnwidth}} {$r/i_{\rm APASS}$ and $r/i_{\rm PS1}$} are treated as the same in SED fitting \tablenotemark{$1$}. If a star has both $r_{\rm PS1}$ (or $i_{\rm PS1}$) and $r_{\rm APASS}$ (or $i_{\rm APASS}$) data, then Pan-STARRS $r$ (or $i$) will be used. \end{tabular} \\
\enddata
\tablenotetext{1}{{Discussed in Section \ref{sec:data}}}
\tablecomments{A star will be included only if it passes both \textit{Gaia} and 2MASS criteria, because the SED fitting method needs at least 4 pass-bands (Section \ref{subsec:fit_photospheric_SED}) and \textit{Gaia} and 2MASS pass-bands cover the full temperature range of Pleiades stars (Section \ref{subsec:selection_Pleiades}, Figure \ref{fig:color_color_used_in_fit} and Table \ref{table:color_color_table}). If a star passes both \textit{Gaia} and 2MASS criteria and fails in WISE/Pan-STARRS/APASS criteria, it will be included and its WISE/Pan-STARRS/APASS photometry will not be used.}
\end{deluxetable*}

A comprehensive description of the star formation history of the Serpens Molecular Cloud requires estimates of when these member stars formed.  However, accurate ages depend on estimates for the stellar properties, which are complicated by the variation of extinction across the cluster.  Cluster ages can be estimated from HR diagrams \citep[e.g.][]{Erickson2015}, the lithium depletion boundary \citep[e.g.][]{Basri1996,binks14}, and the X-ray luminosity function \citep[e.g.][]{Getman2014a}, with varying degrees of accuracy and systematic precision \citep[see review by][]{soderblom14}. However, { for the most common method, use of HR diagrams,} few young stellar objects in the Serpens Molecular Cloud have measured temperatures and luminosities that could be used to estimate their age \citep[e.g.][]{Oliveira2013,Erickson2015}.  In the absence of spectroscopic measurements, photospheric properties are usually estimated by fitting SEDs with synthetic spectra \citep[e.g.,][]{Bayo2008, Robitaille2017, Davies2020}.
This approach may introduce biases at cool temperaturesm where synthetic SEDs differ from observed SEDs (e.g. \citealt{Bell2012} and \citealt{Lancon2020}; see also analysis of main-sequence M-dwarfs in \citealt{mann15}).
Because nearby star-forming regions are mainly composed of low-mass stars \citep[e.g.,][]{Luhman2020}, low-mass stars dominate {isochronal} age measurements for most nearby star-forming regions {\citep[e.g.,][]{Mayne2007,Bell2012,Zari2019}}; any systematic errors in their stellar properties propagate into (often unassessed) uncertainties in ages of star-forming regions.

In this paper, we develop a new SED fitting routine designed for young stars and then apply the fitting routine to optical members of Serpens to estimate stellar ages by comparing best-fit temperature and luminosities to model isochrones. In Section \ref{sec:data}, we present the data, including data for the fitting method and for Serpens. In Section \ref{sec:fitting_method}, details of the SED fitting method are described. In Section \ref{sec:Serpens}, we use the method on Serpens stars and derive their ages. In Section \ref{sec:discussion}, we discuss the results, and the conclusions are summarized in Section \ref{sec:conclusion}. 

\section{Photometric data and cluster member selection}
\label{sec:data}

The results from this paper are obtained by fitting the SEDs of YSOs in Serpens using a grid of empirically measured SEDs of young stars. The grid is built with known members of the Pleiades and nearby young moving groups, following similar approaches by \citet{Bell2012} and \citet{Pecaut2013}.  The Pleiades members are young \citep[$\sim$110 Myr,][]{Gaia2018b}, nearby \citep[$\sim$136 pc,][]{Gaia2018b}, have solar metallicity ($[Fe/H] \sim 0.03$, \citealp{Soderblom2009}), and for the cooler stars are magnetically active (see the effect of surface gravity in Section \ref{subsec:method_problems}). The line-of-sight extinction to the Pleiades is negligible  \citep[$A_V=0.12$,][]{Stauffer1998}, so the photometry does not require any significant correction. This grid and our fitting methods are then tested using known members of Orion.  The photometry used in this paper is obtained from 2MASS \citep{Cutri2003,2MASS_DOI}, ALLWISE \citep{Cutri2014,ALLWISE_DOI}, Pan-STARRS1 \citep[PS1,][]{Chambers2016}, APASS \citep[the Apache Point Observatory Galactic Evolution Experiment,][]{Henden2015}, and {\textit{Gaia} EDR3 \citep{Gaia2021}}, with quality criteria described in Table~\ref{table:photometry_criteria}.

The Pan-STARRS1 and APASS (SDSS-based) filters $r$, $i$ have slightly different transmission curves. 
\citet{Tonry2012} presents $g_{\rm SDSS}-g_{\rm PS1}$, $r_{\rm SDSS}-r_{\rm PS1}$, and $i_{\rm SDSS}-i_{\rm PS1}$ at $>\sim4000$ K (their Figure 6, 7), showing that the difference in $g$ band is significant, while $r$ and $i$ bands have negligible difference. At $<\sim4000$ K, stars in \citet{kado2016} 
 $r_{\rm APASS}$ is fainter than $r_{\rm PS1}$ by $\sim0.07$ mag and $i_{\rm APASS}$ is fainter than $i_{\rm PS1}$ by $\sim0.05$ mag.
In addition, the APASS faint completeness limit (V=16) corresponds to $\sim3700$ K for Pleiades, so this difference is not relevant for most stars. For example, among the selected Pleiades stars under 4000 K (178 stars), only 16 stars have APASS data, and all the 16 stars have Pan-STARRS data. Among the Orion stars used to test the SED fitting method, only 1 star under 4000 K has APASS data, indicating that the magnitude difference has a minor effect on the fitting method.
In SED fitting, we use $g_{\rm APASS}$ and $g_{\rm PS1}$ separately, and treat $r$ and $i$ as the same. If a star has both $r_{\rm PS1}$ (or $i_{\rm PS1}$) and $r_{\rm APASS}$ (or $i_{\rm APASS}$) data, then Pan-STARRS $r$ (or $i$) will be used.

\begin{figure}[!t]
    \centering
    \includegraphics[width=0.9\columnwidth]{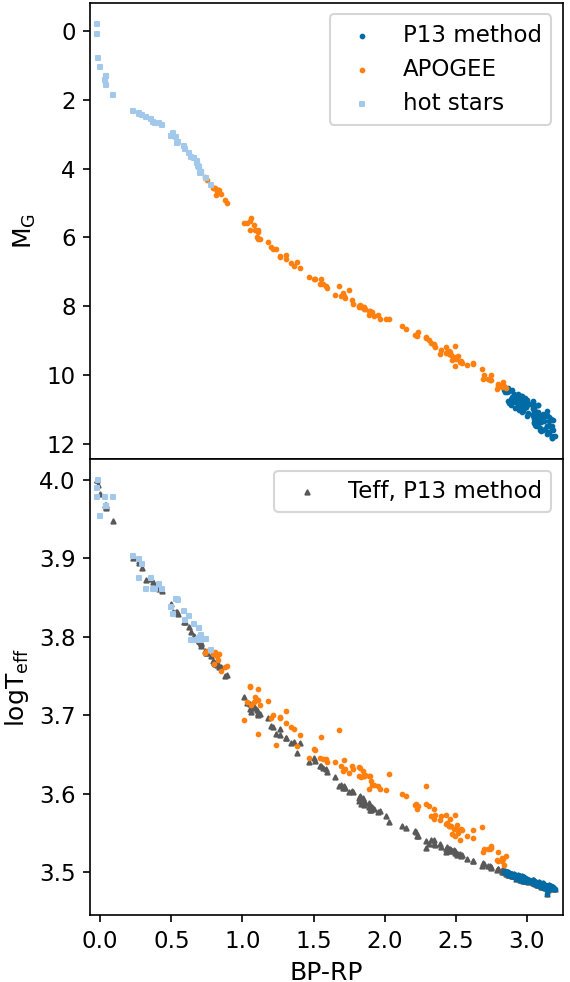}
    \caption{Color-magnitude diagram (top) and color-temperature relation (bottom) of the Pleiades stars in our sample, for cool stars with temperatures derived from \citet{Pecaut2013} SED fits (deep blue dots), stars with APOGEE temperatures (orange dots), hotter stars with temperatures from other catalogs (light blue square), and (in the bottom panel) stars with APOGEE temperatures that also have temperatures assessed by \citet{Pecaut2013} (gray triangle).}
    \label{fig:CMD_and_colorT}
\end{figure}

\begin{figure*}[ht]
    \centering
    \includegraphics[width=\textwidth]{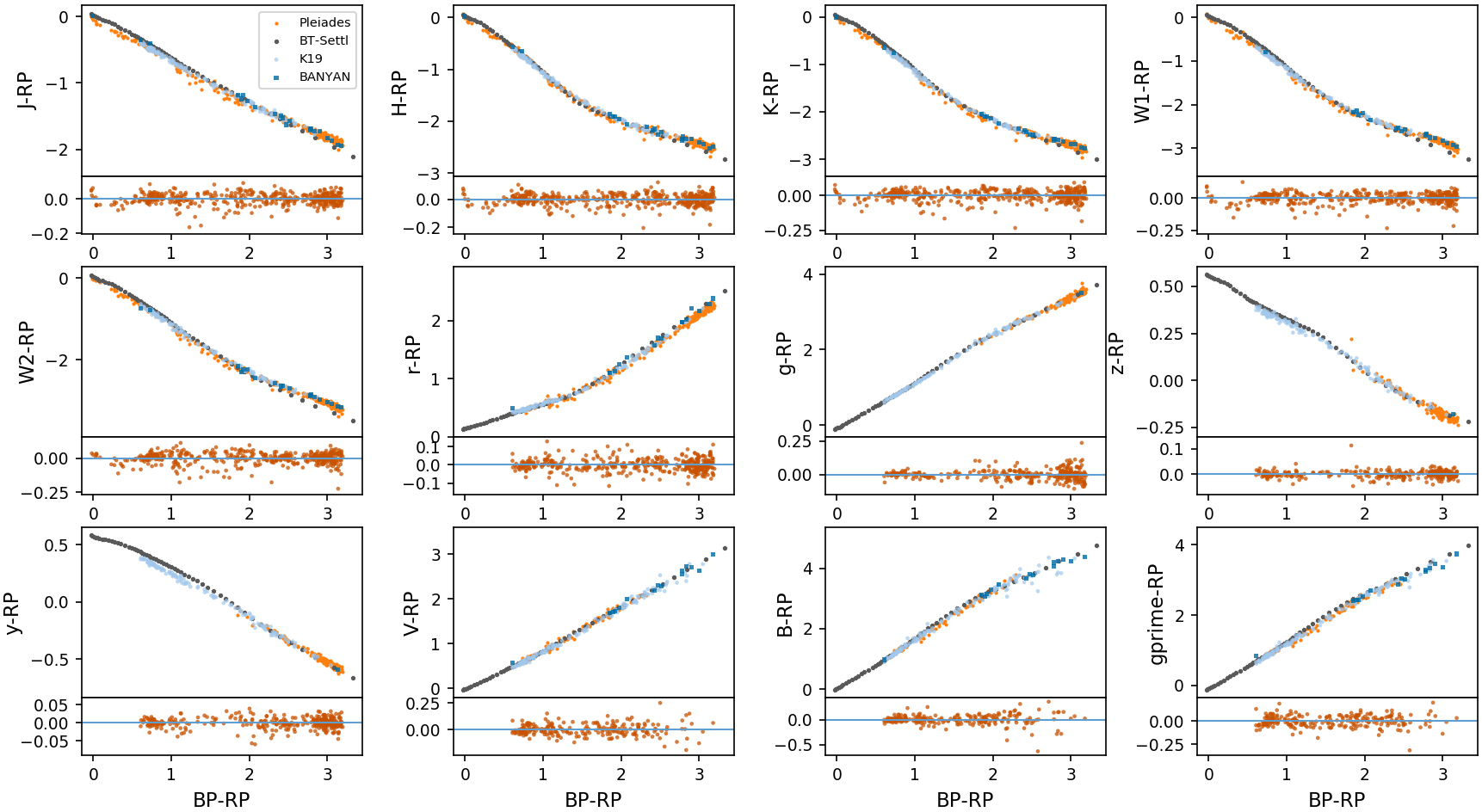}
    \caption{Color-color relations used in SED fitting (except $i$-band) for the set of {Pleiades stars (orange dots), and young stars from \citet{Kounkel2019} (light blue dots) and BANYAN \citep{Gagne2018, Gagne2018b} (deep blue square)}.  Each relationship is shown in the color-color space (top panel) and the residual from a best-fit polynomial. The top panel includes a comparison with colors from BT-Settl spectra with $\log g=4.5$ (gray dots).}
    \label{fig:color_color_used_in_fit}
\end{figure*}

\begin{deluxetable}{lcc}[!bh]
\label{table:Orion_disk_info}
\tablecaption{Selected catalogs of Orion Stars}
\tablehead{\nocolhead{Catalog} & \colhead{Region} & \colhead{Disk ID \tablenotemark{$1$}} }
\startdata
\citet{Fang2013} & L1640 & Yes \\
\citet{Fang2017} & NGC 1980 & Yes \\
\citet{Getman2017} & Orion A& Partial  \\
 & Orion B & Partial\\
 & ONC & Partial\\
 & Flame Nebula & Partial\\
\citet{Kuhn2013} & Flame Nebula & No \\
\citet{Megeath2012} & Orion A, B & No \\
\citet{Prisinzano2008} & ONC & Yes \\
\citet{Kounkel2019} \tablenotemark{$2$} & Orion A, B & Yes \\
& Orion A, B & Yes \\
 & $\sigma$ Ori & Yes \\
 & $\lambda$ Ori & Yes \\
 & 25 Ori & Yes \\
\enddata
\tablenotetext{1}{Disk identification criteria differ among catalogs. When no disk information is provided, spectral indices $\alpha$ are calculated from {\it Spitzer} photometry.  Stars with $\alpha<-1.6$ are treated as diskless stars, following Equation 1 and Section 3.2 in \citet{Dunham2015}.}
\tablenotetext{2}{Targets obtained from \citet{Megeath2012} for Orion A and B, \citet{Hernandez2007} for $\sigma$ Ori, \citet{Suarez2017} for 25 Ori, \citet{Hernandez2010} for $\lambda$ Ori.}
\end{deluxetable}

\subsection{Building a Grid of SEDs of young stars}
\label{subsec:selection_Pleiades}

The photometric grid of young stars is built from Pleiades photometry. The samples that are described here cover from $BP-RP=3.2$ to $0$, corresponding to 3000 K to 9500 K. The cool end is sufficient for fitting bright brown dwarfs in Serpens.

The Pleiades members used for building grids are obtained from several catalogs. First, some Pleiades stars are collected from \citet{Cottaar2014}, which focused on targets with temperatures of 3200--6000 K, as measured from H-band APOGEE spectra. For stars hotter than 6000 K, membership and temperatures are obtained from previous SED fits \citep{Bouy2015,Barrado2016,Somers2017,Bouvier2018}; any hot star in our grid appears in at least two of those catalogs with an adopted temperature averaged from those catalogs.
Stars cooler than 3200 K are identified from \citet{Gaia2018b} and assessed a temperature from the SED fitting method in \citet{Pecaut2013}, who used synthetic spectra to fit the SEDs of stars with negligible extinction. From these catalogs, we then ensure high confidence in membership by requiring that all stars are contained in a 5 pc sphere around the center of the full Pleiades population, in addition to proper motion constraints applied by \citet{Gaia2018b}.  The few Pleiades candidates with $BP-RP>3.2$ are faint and have large uncertainties in $BP$ photometry and are therefore excluded. 

To obtain tight color-color relations, a polynomial is fit to the $BP-RP$ and $G-RP$ relationship. Stars with colors that are more than $2\sigma$ from the fit are removed. We iterate this process until the RMS of the $BP-RP$ and $G-RP$ relation is {$<0.02$ mag}.  A similar procedure is applied to the relationship between $BP-RP$ and $M_G$. These stars form our initial grid for Pleiades (see color-magnitude and color-color diagrams in Figures \ref{fig:CMD_and_colorT}-- \ref{fig:color_color_used_in_fit}).

Due to saturation limits, only Pleiades stars with $T_{\rm{eff}}<4000$ K have Pan-STARRS photometry, while only those with 4000 K $<T_{\rm{eff}}<$ 6000 K have APASS photometry. To supplement our grid, we add young ($<300$ Myr old) stars within 1000 pc, collected from {\citet{KounkelCovey2019}, \citet{Gagne2018}, and \citet{Gagne2018b}.}  These samples are all expected to have roughly solar metallicities \citep{spina17}.  To ensure negligible extinction and the lack of a circumstellar disk, {these additional members are fitted only from Pleiades via the same method in Section \ref{sec:fitting_method} and are required to have:  $| (BP-RP)_{\rm fit} - (BP-RP)_{\rm observed} | < 0.06$, $\chi^2(\rm fit)<0.5$, and $A_V(\rm fit)<0.03$. The fitted result is presented in Appendix \ref{appendix:fit_additional_YSOs}}. This total sample provides Pan-STARRS photometry for stars cooler than 7000 K, sufficiently covering the temperature range of most young stars.

\subsection{Stars in Orion to test and apply the SED fits}
\label{subsubsec:selection_Orion}

To test the accuracy of our fitting methods, we apply our SED fits to stars in the Orion Star-forming Complex, leveraging the photometric and spectroscopic analyses by various catalogs (Table \ref{table:Orion_disk_info}) that include an evaluation of disks from {\it Spitzer} mid-IR photometry.\footnote{Catalogs disagree on disk presence for $\sim 10$ stars because of differences in disk identification methods.  A star is treated as diskless only if all catalogs in which the star is listed identify it as diskless.} Temperatures are obtained from APOGEE\footnote{Some methodological differences still exist between APOGEE measurements because of differences in synthetic models and abundance assumptions used in different papers, see for example temperature measurements of the same set of stars in \citet{DaRio2016} and \citet{Kounkel2018}. {The APOGEE $T_{\rm eff}$ in \citet{Cottaar2014} and \citet{Kounkel2018} is measured from BT-Settl and PHOENIX grid separately.}} spectra \citep{Kounkel2018}. Samples from catalogs in Table \ref{table:Orion_disk_info} are cross-matched with \citet{Kounkel2018} to ensure astrometric membership in Orion. Samples from \citet{Fang2013, Fang2017} are used to identify stars with and without disks (Section \ref{subsec:identify_disk}).

\subsection{The sample of Serpens stars}
\label{subsubsec:selection_Serpens}
The SED fits are designed to derive the stellar properties of the $\sim700$ high-confidence members\footnote{The statistical membership is $\sim 1167$ optical members, which includes thousands of possible members with low confidence in membership.  The stars selected for this study have high confidence of membership based on astrometry.} of the Serpens star-forming complex with \textit{Gaia} DR2 astrometry \citep{Herczeg2019}. 
About half of the optical members are located in three distinct clusters, named Serpens Main ($\sim 440$ pc), Serpens Northeast (Serpens NE, $\sim 480$ pc) and Serpens far-South ($\sim380$ pc). Of these clusters, Serpens Main is rich in protostars \citep{Dunham2015}, while the others have few protostars and disks.  In addition, the Serpens South cluster is rich in protostars but has few optical counterparts.
A few optical members of the Serpens cloud are located in smaller groups, and the rest are distributed throughout the complex. Among the $\sim700$ high confidence members in \citet{Herczeg2019}, $\sim20$ stars have no \textit{Gaia} $BP$ or $RP$ photometry, $\sim20$ stars have no 2MASS data with a crossmatch radius of 2 arcsec, and $\sim80$ stars do not pass the 2MASS data selection criteria, in most cases because the objects appear elongated by a companion or visual binary.  Thus, over 100 stars are removed from the sample, leaving 569 stars for analysis. 

\subsection{Synthetic Spectra and filter profiles}

Synthetic spectra are used for comparisons to observed color-color relations (e.g. Section \ref{subsec:selection_Pleiades}) and for estimating the effect of surface gravity and circumstellar disk in SED fitting (Section \ref{subsec:method_problems}). Synthetic spectra are obtained from BT-Settl models \citet{Allard2012} with solar abundances from \citet{Asplund2009}. {Filter profiles and zero points are collected from the SVO Filter Profile Service \citep{SVO2012,SVO2020} for the following filter sets: GAIA/GAIA3 (GAIA eDR3 release), 2MASS/2MASS, WISE/WISE, PAN-STARRS/PS1, and Misc/APASS.} The synthetic spectra from model atmospheres are used to assess the effects of surface gravity (see {Section \ref{subsubsec:Ple_Ori_color_difference}}).
More broadly, differences between the Pleiades colors and synthetic colors from the model spectra, as seen in Figure \ref{fig:color_color_used_in_fit}, demonstrate why the SED fitting requires empirical colors.

\subsection{Pre-main-sequence Evolutionary Models}
To estimate ages, we adopt the \citet{Feiden2016} non-magnetic evolutionary models. For comparison, the magnetic isochrones from \citet{Feiden2016}  and the PARSEC isochrones \citep{Bressan2012, ChenYang2014, ChenYang2015, TangJing2014} are also used to evaluate uncertainties related to different models of pre-main sequence evolution.

\section{Developing an SED fitting method}
\label{sec:fitting_method}

\subsection{Overview}
\label{subsec:overview}
In our fitting method, we first create SEDs from the combination of temperature $T_{\rm{eff}}$, luminosity $L$, and extinction $A_V$ (Section \ref{subsec:generate_SED}) and then compare these SEDs with the observed stellar SED (Section \ref{subsec:fit_photospheric_SED}). Since YSOs may contain a dusty accretion disk that affects the observed SED, we separately treat disk and diskless stars (Section~\ref{subsec:identify_disk}). The fitting method is tested in Section \ref{subsec:method_test} and compared with other fitting methods in Section \ref{subsec:compare_other_fitting_method}. Caveats of this method are discussed in Section \ref{subsec:method_problems}.

\subsection{Generating SEDs}
\label{subsec:generate_SED}
Pleiades stars are used to generate empirical SEDs with photometry in the following filters:  $BP, RP, J, H, K, W1, W2, g, r, i, z, y, g', B, V$. Figure \ref{fig:CMD_and_colorT} and \ref{fig:color_color_used_in_fit} show relationships between colors, magnitudes, and temperatures for the Pleiades stars and young stars from other catalogs (see also additional color relationships in Appendix \ref{appendix:sec:other_color_relations}). In general, young stars from these catalogs are consistent with Pleiades stars in color-color relations. Colors from BT-Settl spectra differ from Pleiades relations, especially at temperatures cooler than $\sim4000$ K, as expected from previous assessments \citep[e.g.][]{Bell2012, Herczeg2015,Lancon2020}. 

First, colors in Figure \ref{fig:color_color_used_in_fit} are reddened by {visual extinction $A_V$ and relative extinction $A_X/A_V$}, as obtained from the analysis of red clump stars with total-to-selective extinction coefficient $R_V=3.16$ \citep{Wang2019} {(except $BP$ and $RP$)}, {and from \citet{Gange2020} for Gaia $BP$ and $RP$}. 
Extinction corrections are applied only for the \textit{Gaia} $BP$ and $RP$ photometry, based on the temperature dependence determined by \citet{Gange2020}\footnote{{\citet{Gange2020} uses \citet{Fitzpatrick1999} extinction law with $R_V=3.1$ to calculate \textit{Gaia} extinction. The \citet{Wang2019} extinction law is consistent with \citet{Fitzpatrick1999} in the optical and is steeper in the infrared. The extinction of both $BP$ and $RP$ in \citet{Wang2019} corresponds to \citet{Gange2020} extinction at $\sim5300$ K.}}. Then, the reddened colors are added by $m_{RP}$, leading to the generated SEDs for use in our fits.
 
The extinction-corrected color $BP-RP$ from the best fit is converted to $T_{\rm eff}$ from the color-temperature relationship established from Pleiades stars (see the bottom panel of Figure \ref{fig:CMD_and_colorT}). The APOGEE $T_{\rm eff}$ is preferentially adopted when available, despite systematic differences between the APOGEE scale and the color-temperature relationship of \citet{Pecaut2013}, as shown in Section \ref{subsec:selection_Pleiades} and the bottom panel of Figure \ref{fig:CMD_and_colorT} where the APOGEE $T_{\rm eff}$ of Pleiades stars only covers 3200-6000 K and differs from \citet{Pecaut2013} $T_{\rm eff}$. The SED fit also solves for an apparent magnitude $m_{RP}$, which is then converted to luminosity using bolometric corrections (Section \ref{subsec:fit_photospheric_SED}), the extinction, and the distance.  The temperatures in our fits range from 3000--9500 K and extinctions range from $0<A_V<6$.

The broadband SEDs generated from the empirical color-color relations are then fit to the photometry (in mag) of each star.  

\begin{figure}[!t]
    \centering
    \includegraphics[width=\columnwidth]{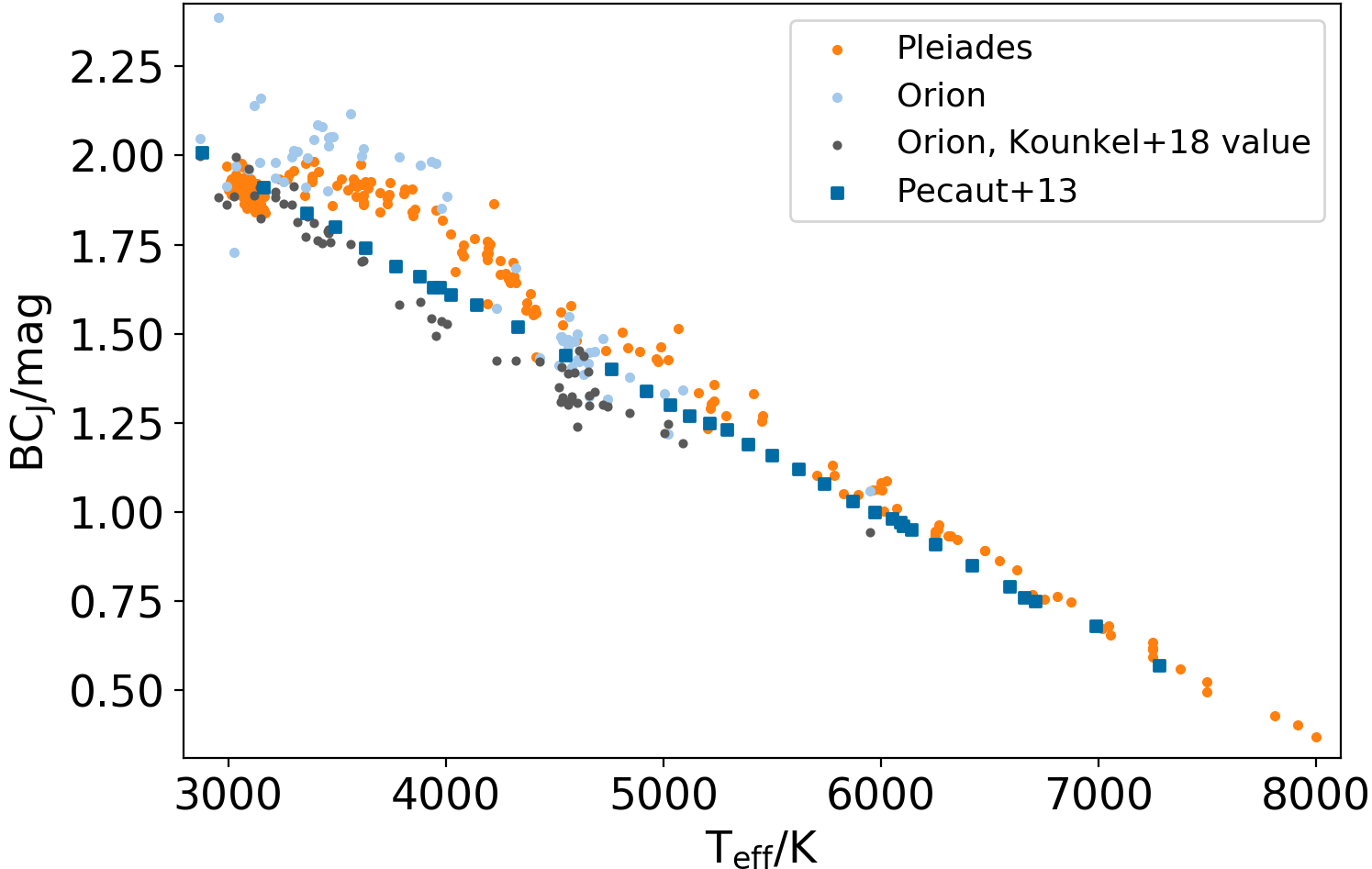}
    \caption{{The relationship between the $J$-band bolometric correction and temperature directly from \citet{Pecaut2013} in deep blue, the Pleiades APOGEE temperatures with bolometric corrections from \citet{Pecaut2013} in orange, the Orion APOGEE temperatures with the \citet{Pecaut2013} bolometric correction in light blue, and 
    Orion APOGEE temperature with $BC_J$ converted from \citet{Kounkel2018} bolometric fluxes in gray.)}}
    \label{fig:BCJ_comparison}
\end{figure}

\subsection{Fitting photospheric SEDs}
\label{subsec:fit_photospheric_SED}
In our fitting procedure, we find the minimum reduced $\chi_{\rm red}^2=\chi^2 / (N-3)$, where $N$ is the number of data points, using the function \texttt{scipy.optimize.minimize} in python \citep{Virtanen2019}. The free parameters in the fit are an extinction-corrected $BP-RP$ (temperature), $A_V$ (extinction), and $m_{RP}$ (visual magnitude, converted to luminosity after extinction and bolometric corrections). Since these parameters have some degeneracies, the ``minimize" function starts at 15 equally spaced locations of $BP-RP$ and 10 equally spaced locations of $A_V$. 

{ For each photometric band, the uncertainty is estimated from a global evaluation of the average standard between the best-fit model and empirical measurement.  The photometric uncertainty is {first} assumed to be $0.1$ mag to account for variability, unless the listed empirical error is larger than 0.1 mag.  After fitting to all targets, the magnitude difference in all pass-bands is calculated.  The standard deviation for each filter is then adopted as the magnitude uncertainty in the subsequent fitting process.  We iterate one additional sequence to adopt final empirical uncertainties and measurements.}  Errors of the fitted parameters are calculated based on {fits} with $\chi^2 - \chi^2_{\rm min} < 3$ \citep{Robitaille2007}.  Best-fit results at the edge of the fitting range are included in our tests (Section \ref{subsec:method_test}) but are excluded from any final results of Serpens members (Section \ref{subsec:Serpens_ages}).

The fits use as many of the photometric points as possible. However, as shown in Figure \ref{fig:color_color_used_in_fit} (or Table \ref{table:color_color_table}) and described in Section \ref{subsec:selection_Pleiades}, some photometry is only available over a limited range of $BP-RP$. For example, Pan-STARRS $g$ band photometry only exists at $0.8<BP-RP<3.2$, while APASS $V$ band photometry is available at $0.03<BP-RP<3$. Thus, when calculating $\chi^2$, $N$ will change with the input $BP-RP$, and a pass-band $X$ will not be used if the input $BP-RP$ exceeds the available $BP-RP$ range of pass-band $X$. Under these circumstances, a new minimum $\chi^2$ will be calculated for the relevant pass-bands in the previously fitted $BP-RP$. \footnote{As an example, consider a star with observed photometry of $BP, RP, J, H, K, g$.  Because our grid includes Pan-STARRS $g$ only for $0.8<BP-RP<3.2$, the minimum $\chi^2$ will include $g$ at $BP-RP>0.8$ and exclude $g$ at $BP-RP<0.8$, changing the number of data points $N$ that lead to the $\chi^2$.  
If the resulting minimum $\chi^2$ is located at $BP-RP>0.8$, it will be the final minimum $\chi^2$. Otherwise, a new minimum $\chi^2$ will be calculated by excluding $g$ in full $BP-RP$ range.}

The extinction-corrected $BP-RP$ of the best-fit SED is converted to temperature using the relationship determined from the Pleiades. 

\subsection{Bolometric corrections}
\label{subsec:bolometric_correction}

The luminosity\footnote{$M_{bol\odot}=4.74$ as defined by the IAU Resolution 2015 at https://www.iau.org/static/resolutions/IAU2015$\_$English.pdf.} is calculated from $m_{RP}$, corrected for extinction and a $J$ band bolometric correction ($BC_J$) obtained for the relevant temperature from \citet{Pecaut2013} for stars cooler than 7300 K. For hotter stars, the bolometric correction from $J$ band is obtained from PARSEC isochrones.

Since different $T_{\rm eff}$ calculation methods affect $BC_J$-$T_{\rm eff}$ relation, we compare $BC_J$-$T_{\rm eff}$ relations calculated from \citet{Pecaut2013} $T_{\rm eff}$ and APOGEE $T_{\rm eff}$. 
For these comparisons, we  with APOGEE $T_{\rm eff}$ include Pleiades stars (described in Section \ref{subsec:selection_Pleiades}) and Orion stars (described in Section \ref{subsubsec:selection_Orion}). The Orion stars are diskless stars with {$A_V<=0.5$}\footnote{{To avoid extinction issue, stars with negligible extinction are best. Because no stars have $A_V=0$ and only 21 stars have $A_V<=0.2$, we loosen the constraint to $A_V<=0.5$.}} and luminosities calculated by \citet{Kounkel2018}). Bolometric correction of these Orion stars via \citet{Kounkel2018} method is also included, where they derived bolometric flux from SED fitting and APOGEE $T_{\rm eff}$. The result ($BC_J$-$T_{\rm eff}$ relation) is shown in Figure \ref{fig:BCJ_comparison}, with the following results.
When using APOGEE $T_{\rm eff}$ and \citet{Pecaut2013} bolometric correction method,  Pleiades and Orion relations overlap, with deviations at {$\sim4000$ K}.
Some of this difference maybe be caused by the discrepancy between APOGEE $T_{\rm eff}$ and \citet{Pecaut2013} temperatures, as seen in the bottom panel of Figure \ref{fig:CMD_and_colorT}.

\subsection{Identifying stars with disks}
\label{subsec:identify_disk}

\begin{figure}[bht]
    \centering
    \includegraphics[width=\columnwidth]{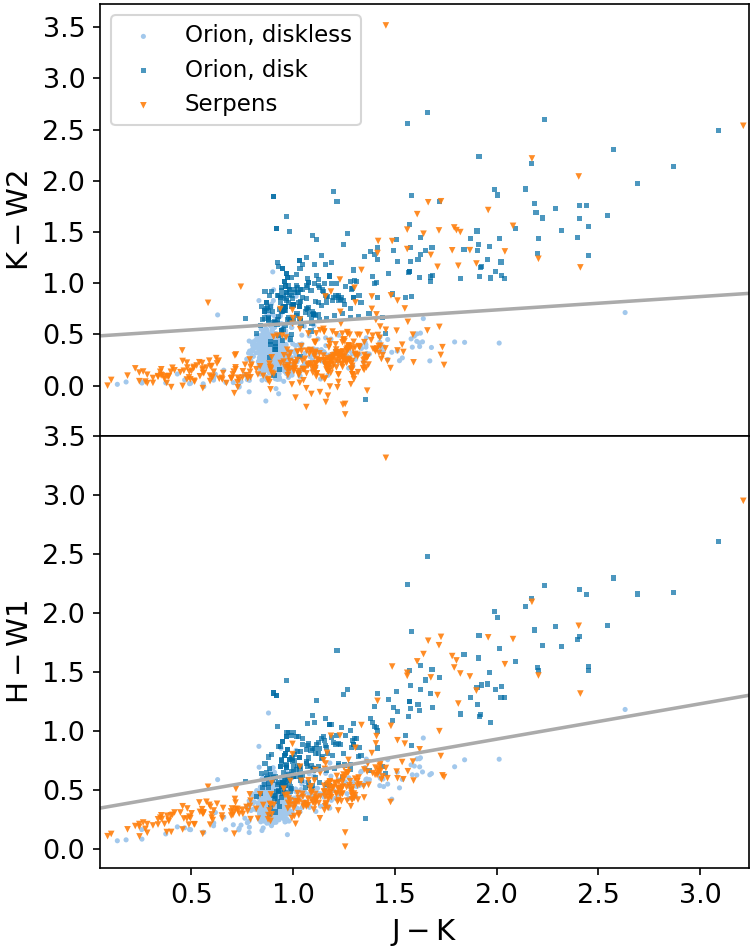}
    \caption{Color-color relationships to separate disk-hosting and diskless stars.  In Orion, stars with disks (deep blue) and without disks (light blue) based on {\it Spitzer} photometry, separated by the solid gray line.  Most optical members of Serpens (orange) are diskless stars.}
            \label{fig:disk_selection_KW2_HW1}
\end{figure}

\begin{figure*}[ht]
    \centering
    \includegraphics[width=\textwidth]{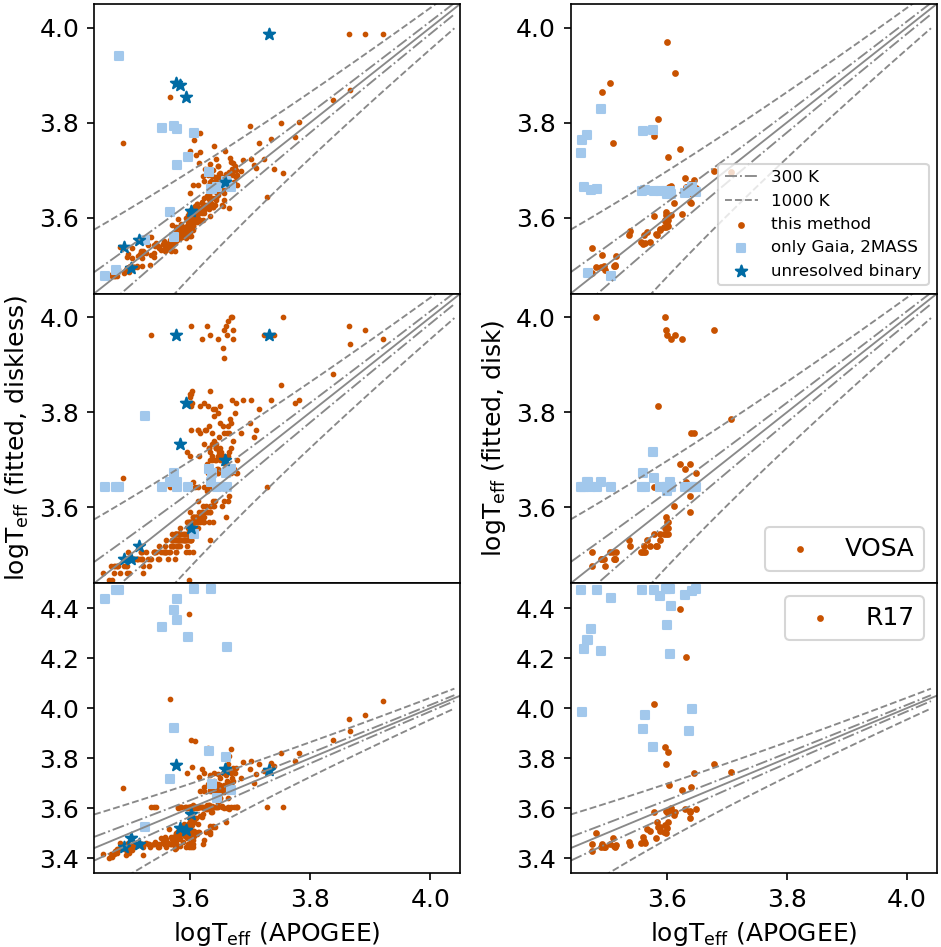}
    \caption{Comparison between temperatures from SED fits and literature (APOGEE) temperatures for diskless (left) and disk (right) stars in Orion, for our SED fitting method (top), VOSA \citep[middle,][]{Bayo2008}, and \citet{Robitaille2017} (bottom). 
    The gray solid line represents equal temperature line, while gray dashed lines represent deviation of 300 K and 1000 K. The light blue square symbols are stars with only \textit{Gaia} and 2MASS data in the fitting. The deep blue asterisk symbols are stars with a nearby source resolved by \textit{Gaia} and within 2 mag in $G$ band, but unresolved by 2MASS.}
    \label{fig:fitting_result_compare}
\end{figure*}

\begin{figure}[!t]
    \centering
    \includegraphics[width=\columnwidth]{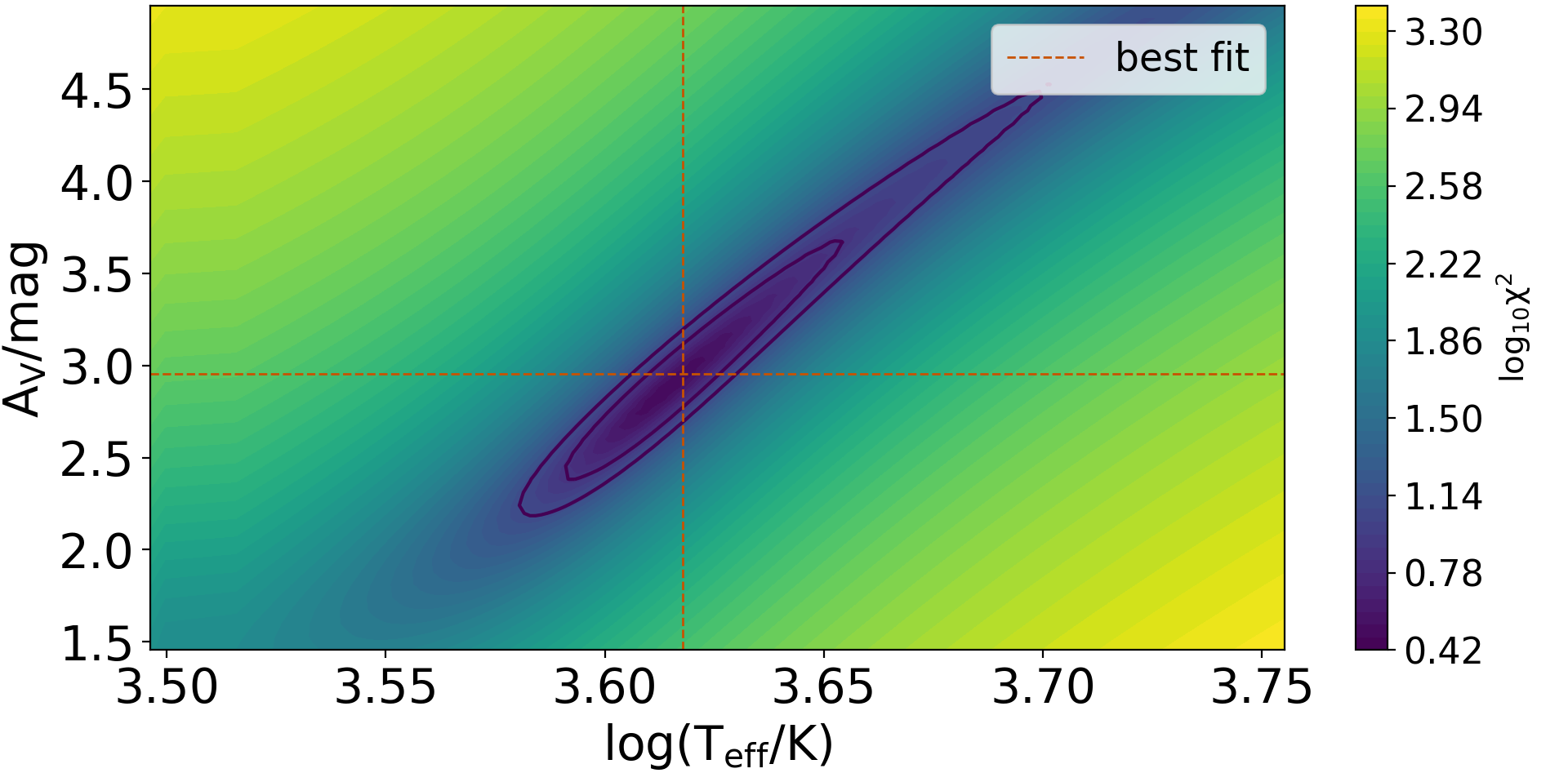}
    \caption{{Contour of a diskless Orion star in the fitting calculated from grids of $T_{\rm eff}$ and $A_V$. Color bar represents $\log_{10}(\chi^2)$. The black contours are $\chi^2-\chi^2_{min}=5$ and $\chi^2-\chi^2_{min}=10$. The orange lines are the best-fit $T_{\rm eff}$ and $A_V$.}}
    \label{fig:chi2_contour}
\end{figure}

The presence of a disk introduces complications to interpreting SEDs. Warm dust in accretion disks produce excess IR emission, especially in the all-sky WISE survey \citep[see review by][]{Williams2011}.  Accretion produces excess H continuum emission that makes optical colors bluer, in some cases dominating over any photospheric emission \citep[see review by][]{hartmann16}. For some viewing angles, the disk intercepts our line-of-sight to the star, so the observed emission from the star is extinguished and contaminated by scattered light. These processes are all variable \citep[e.g.][]{grankin07,guo18gitau,venuti21}, which introduces additional uncertainties when comparing non-simultaneous data.

Figure \ref{fig:disk_selection_KW2_HW1} shows the 2MASS and WISE color-color relations used to separate disk and diskless stars, based on the identification of disks in the Orion sample from excess emission in {\it Spitzer} mid-IR imaging \citep{Fang2013,Fang2017}. To quantify the separation (Figure~\ref{fig:disk_selection_KW2_HW1}), we follow the ``Class II box" method in \citet{Dunham2015} with disk and diskless stars separated by a line that  maximizes the sum of the two kinds of probability: (a) the probability that a diskless star is located under the line (defined as the number of diskless stars under the line divided by the total number of diskless stars), and (b) the probability that a star under the line is classified as diskless (defined as the number of diskless stars under the line divided by the total number of stars under the line). Disks are identified from excess emission at $W2$, as seen in the $K-W2$ (or $H-W1$, if $W2$ photometry is unreliable) versus $J-K$. For stars with disks, $W1$ and $W2$ photometry are excluded from SED fits.

\subsection{Testing the SED fitting results}
\label{subsec:method_test}

To test the fitting method, we fit the SEDs of YSOs in Orion and then compare the resulting temperatures to those measured from APOGEE spectra \citep{Cottaar2014,Kounkel2018}. Fits to stars with and without disks are analyzed separately.  We exclude stars that only have \textit{Gaia} and 2MASS photometry and visual binaries that are resolved by Gaia and unresolved by 2MASS.

Figure~\ref{fig:fitting_result_compare} and Table~\ref{table:fitting_deviation} show the difference and scatter of fitted temperatures \footnote{Represented by mean and standard deviation of $(T_{\rm eff, fit}-T_{\rm eff, APOGEE})$}. In general, the difference and scatter of $T_{\rm eff}$ are 90 K and 433 K for stars without disks and 533 K and 1246 K for stars with disks. Our method is more reliable for both diskless and disk-hosting stars with $T_{\rm fit}< \sim$5000 K, where the difference and scatter are only -13 K and 208 K for diskless stars, and 53 K and 212 K for disk stars. For disk stars with higher fitted temperatures, the fit may be unreliable, affected by variability and by contamination of photospheric emission by emission from the accretion disk. Stars that only have \textit{Gaia} and 2MASS photometry and also visual binaries that are resolved by \textit{Gaia} and unresolved by 2MASS are included in Figure~\ref{fig:fitting_result_compare} to demonstrate that our method is not reliable for these stars.  These stars are excluded in calculating the average temperature differences listed above.

\begin{figure}[!t]
    \centering
    \includegraphics[width=0.9\columnwidth]{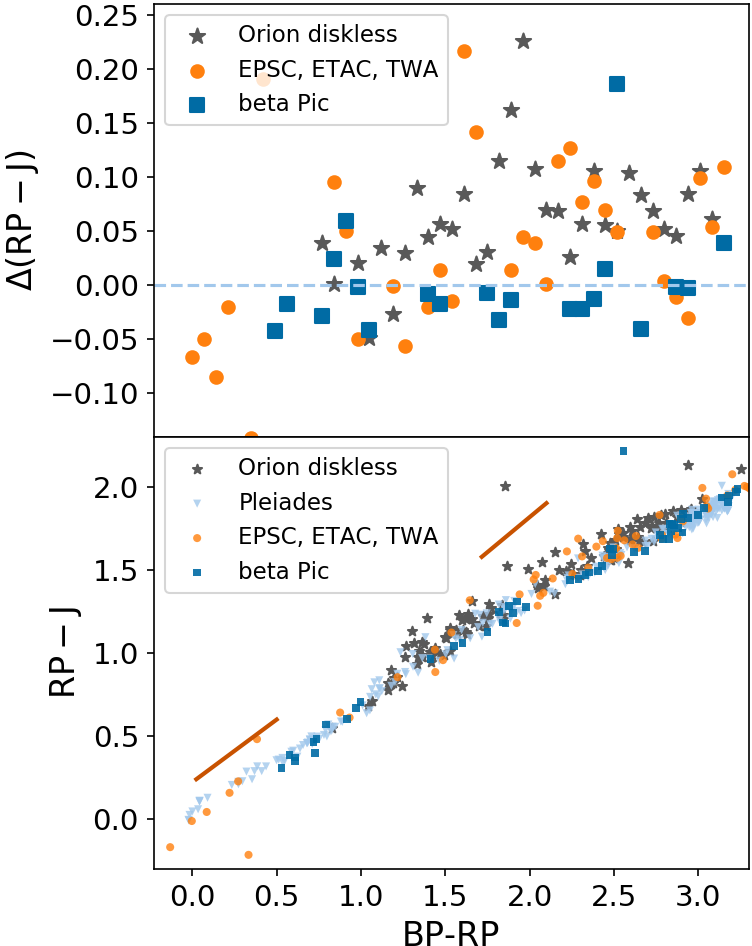}
    \caption{Color color diagrams (bottom) and residuals (top) for stars in groups with different ages.  The Pleiades stars (light blue triangles) form the foundation for our fits, and are compared with diskless stars in Orion with $A_V<1$ (gray asterisk), the beta Pic Moving Group (deep blue square),  $\epsilon$ Chamaeleontis, $\eta$ Chamaeleontis and TW Hya Association (orange). Deep orange line segments in the bottom panels are extinction vectors ($A_V=1$) at $BP-RP=0.5$ (left) and $BP-RP=2.1$ (right).}
    \label{fig:quantify_RP_J}
\end{figure}

{Figure~\ref{fig:fitting_result_compare} also shows that there are more stars with $T_{\rm fit}-T_{\rm APOGEE}>300~K$ at $T_{\rm APOGEE}>\sim4500~K\ (\sim3.65$~dex) than at cooler $T_{\rm APOGEE}$, which is likely due to the larger degeneracy between $T_{\rm eff}$ and $A_V$ at higher temperature. Figure \ref{fig:chi2_contour} shows the $\chi^2$ contours around the best-fit parameters for a diskless star, as calculated from grids of $T_{\rm eff}$ and $A_V$. The contours extend to high temperatures because of a large degeneracy between temperature and extinction. Such degeneracy is also shown in the bottom panel of Figure \ref{fig:quantify_RP_J}, where the color-color relation is closer to parallel with the extinction vector at higher temperature. }

\begin{figure*}[ht]
    \centering
    \includegraphics[width=0.85\textwidth]{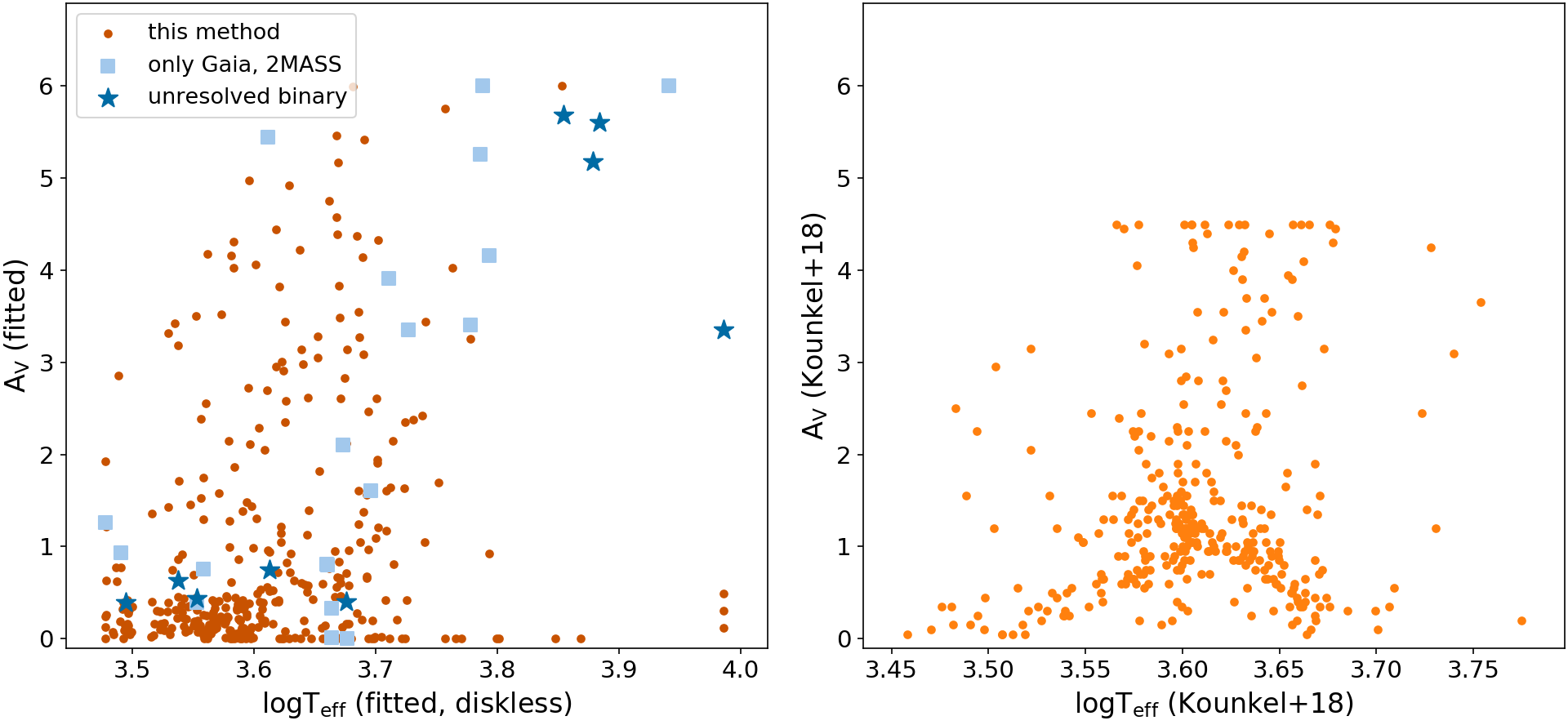}
    \caption{$T_{\rm eff}$ versus $A_V$ for diskless Orion stars, for (left) our results and (right) fits from \citet{Kounkel2018}.  The colors and symbols are the same as those in Figure \ref{fig:fitting_result_compare}.}
    \label{fig:Orion_T_Av_fitted_literature}
\end{figure*}

\begin{deluxetable*}{cccccccc}
\label{table:fitting_deviation}
\caption{Temperature difference and scatter \tablenotemark{1} (in unit of K) between fitted and literature result for stars with the highest data quality \tablenotemark{2}.}
\tablehead{\colhead{$T_{\rm eff, fit}$ range (K)} & & & \colhead{Diskless} & & &  \colhead{Disk}}
\startdata
 & &  this work & VOSA & R17 & this work & VOSA & R17 \\ \hline
\multirow{2}{*}{\textless{}3500} & Difference & -40 & -323 & -690 & 47 & -194 & -475 \\
 & Scatter & 143 & 225 & 245 & 159 & 206 & 210 \\ \hline
\multirow{2}{*}{3500-4000} & Difference & -129 & -369 & -431 & -69 & -344 & -306 \\
 & Scatter & 98 & 105 & 104 & 124 & 117 & 210 \\ \hline
\multirow{2}{*}{4000-5000} & Difference & 117 & 210 & -21 & 186 & 392 & 605 \\
 & Scatter & 240 & 436 & 389 & 234 & 246 & 199 \\ \hline
\multirow{2}{*}{\textgreater{}5000} & Difference & 857 & 2050 & 1565 & 3092 & 4114 & 5524 \\
 & Scatter & 779 & 1474 & 2795 & 1350 & 1932 & 6366 \\ \hline
\multirow{2}{*}{All} & Difference & 90 & 380 & -109 & 533 & 774 & 571 \\
 & Scatter & 433 & 1324 & 1345 & 1246 & 1961 & 3331 \\
\enddata
\tablenotetext{1}{Difference and scatter are the mean and standard deviation of $(T_{\rm fit}-T_{\rm APOGEE})$ separately.}
\tablenotetext{2}{{Stars with only \textit{Gaia} and 2MASS data are excluded; stars with a nearby source unresolved by 2MASS are excluded. The rest are marked as ``highest data quality''.}}
\tablecomments{The columns marked with ``VOSA'' and ``R17'' are comparison of VOSA and \citet{Robitaille2017} SED fitting tools with details described in Section \ref{subsec:method_test} and Section \ref{subsec:compare_other_fitting_method} separately.}
\end{deluxetable*}

\begin{deluxetable*}{ccccccccccccccccc}
\label{table:banduncertainties}
\tablecaption{{Average difference between model and observed magnitude}}
\tablehead{\colhead{Region} & \colhead{disk} & \colhead{$T_{\rm eff, APOGEE}$} & \colhead{$\Delta g$} & \colhead{$\Delta B$} & \colhead{$\Delta BP$} & \colhead{$\Delta V$} & \colhead{$\Delta r$} & \colhead{$\Delta i$} & \colhead{$\Delta RP$} & \colhead{$\Delta z$} & \colhead{$\Delta y$} & \colhead{$\Delta J$} & \colhead{$\Delta H$} & \colhead{$\Delta K$} & \colhead{$\Delta W1$} & \colhead{$\Delta W2$}}
\startdata
  Orion & diskless & $<=4000$ K & 0.13 & 0.3 & 0.02 & 0.10 & 0.08 & 0.03 & 0.03 & 0.01 & 0.02 & 0.03 & 0.03 & 0.02 & 0.03 & 0.04 \\
   &   & $4000-5000$ K & 0.21 & 0.21 & 0.02 & 0.10 & 0.06 & 0.04 & 0.06 & 0.02 & 0.03 & 0.03 & 0.03 & 0.03 & 0.03 & 0.05 \\
    &   & $>5000$ K & 0.16 & 0.20 & 0.01 & 0.08 & 0.05 & 0.05 & 0.03 & 0.01 & 0.03 & 0.04 & 0.02 & 0.02 & 0.03 & 0.05 \\
    &   & all       & 0.16 & 0.23 & 0.02 & 0.10 & 0.07 & 0.04 & 0.04 & 0.01 & 0.02 & 0.03 & 0.03 & 0.02 & 0.03 & 0.05 \\
  Orion & disk & $<=4000$ K & 0.24 & 0.35 & 0.07 & 0.11 & 0.06 & 0.04 & 0.14 & 0.03 & 0.03 & 0.04 & 0.01 & 0.09 & 0.15 & 0.35  \\ 
   &  &  $4000-5000$ K  & 0.47 & 0.21 & 0.07 & 0.07 & 0.18 & 0.08 & 0.17 & 0.05 & 0.05 & 0.08 & 0.04 & 0.15 & 0.15 & 0.34  \\ 
    &  &  $>5000$ K  & - & - & - & - & - & - & - & - & - & - & - & - & - & -  \\
    &  &  all  & 0.39 & 0.22 & 0.07 & 0.07 & 0.10 & 0.05 & 0.15 & 0.04 & 0.03 & 0.05 & 0.02 & 0.10 & 0.15 & 0.34 \\
\enddata
\tablecomments{Although the fitting on disk stars excludes $W1$ and $W2$, these pass-bands are still included in the table, representing the difference between model stellar photosphere SEDs and observed SEDs with infrared excess.}
\tablecomments{{Only 1 disk star has $T_{\rm eff,APOGEE}>$5000 K, thus we do not report its magnitude difference.}}
\end{deluxetable*}

Figure \ref{fig:Orion_T_Av_fitted_literature} shows the relationship between extinction and $T_{\rm eff}$ for diskless Orion stars from our results and from \citet{Kounkel2018}. The distribution of extinctions in \citet{Kounkel2018} is $\sim 1$ mag higher for stars with $\log (T/K)\sim3.6$ than for hotter and cooler stars, possibly due to differences between observed SEDs and synthetic spectra. \citet{Kounkel2018} first determined temperature from APOGEE spectra and then compared the observed SED with PHOENIX synthetic spectra \citep{Husser2013} with the same $T_{\rm eff}$ to calculate $A_V$. If the stellar photosphere is redder than the synthetic spectrum, the calculated $A_V$ will be higher, leading to a gap in $T_{\rm eff}$ - $A_V$ relation (perhaps caused by the discrepancy seen in color-temperature scale in Figure~\ref{fig:CMD_and_colorT}). Such a gap is not seen in our fitting method (Figure \ref{fig:Orion_T_Av_fitted_literature}, left panel), { which is consistent with} Orion stars having colors that are more similar to the Pleiades SEDs than to synthetic spectra.

Table \ref{table:banduncertainties} lists the average difference between fitted and observed magnitude of these Orion stars after two iterations of SED fits. In general, for diskless stars, the average magnitude difference {is less than $\sim0.03$ mag at pass-bands redder than $RP$}. The larger uncertainties in blue wavelengths (except $BP$) are likely introduced by differences in spot coverage and by variability.
For disk stars, the smallest average magnitude difference is at $H$ band. The $W2$ band, which is excluded in these fits, shows the largest difference, as expected for stars with excess infrared emission from disks.  Optical pass-bands $BP$ and $RP$ have larger differences than the near-infrared passbands, likely because accretion variability and disk obscuration affect the optical more than the near-IR.

Additional descriptions {and comments} of these results are provided in {Section \ref{subsec:method_problems}}.

\subsection{Comparing our SED fits to other fitting methods}
\label{subsec:compare_other_fitting_method}

To compare with other SED fitting methods, photometry of stars in Orion are fit with the commonly used SED codes VOSA \citep{Bayo2008} and \citet{Robitaille2017} (hereafter Robitaille).  Our fits with VOSA use the BT-Settl models with solar metallicity and surface gravity $\log (g)=4.5$, an extinction law from \citet{Fitzpatrick1999} and improved by \citet{Indebetouw2005} in the infrared.  The Robitaille fits  use their ``s---s-i" models (pure stellar photosphere models constructed from \citet{Castelli2003} above 4000 K and PHOENIX atmospheres \citep{Brott2005} below 4000 K) with an extinction law consistent with $R_V=5.5$ \citep[following the description in][]{Forbrich2010}.

Figure \ref{fig:fitting_result_compare} and Table \ref{table:fitting_deviation} show that the difference and scatter between the fitted and literature (spectroscopic) temperatures are 380 K and 1324 K for VOSA, -109 K and 1345 K for \citet{Robitaille2017}, and 90 K and 433 K for our method. All three methods are less reliable for stars that have only \textit{Gaia} and 2MASS photometry and visual binaries that are resolved by \textit{Gaia} and unresolved by 2MASS. In addition, at $\log T/K<3.6$, both the VOSA and Robitaille fits underestimate temperatures by {$\sim$300 K to $\sim$1000~K} relative to literature values. Both models also assess a wide range of temperatures for stars with spectroscopic temperatures of $\log T/K\sim 3.65$. In comparison, temperatures from our fits follow the literature estimates within $\sim 300$ K until $\log T/K=3.65$, where our models suffer from a similar but less severe dispersion.  Possible reasons for this feature are discussed in {Section \ref{subsec:method_problems}}.

In these comparisons, the stellar magnitude and fitting ranges are the same as Section \ref{subsec:method_test}. The \citet{Robitaille2017} fits do not include additional temperature constraints, thus the fitted temperature can exceed the ranges of our method. Magnitude uncertainties are 0.1 mag in both VOSA and \citet{Robitaille2017} fits, with no iteration executed in their methods.

\subsection{Shortcomings and challenges in our SED fitting}
\label{subsec:method_problems}

This SED fitting method uses Pleiades stars to create templates to fit YSOs in the Serpens Molecular Clouds.  However, the Pleiades stars are not perfect templates. Shortcomings caused by Pleiades include differences in surface gravity and spots (see Section \ref{subsubsec:Ple_Ori_color_difference}). In addition, we comment on our fits to disk stars in Section \ref{subsubsec:fits_to_disk_stars}, and show how positive extinction affects the fitting result on low extinction clusters (Section \ref{subsubsec:constrained_extinction}). Finally, we comment the outliers of the fitting result in Section \ref{subsubsec:outliers_Teff_comparison}.

\subsubsection{Temporal evolution of color-color relations}
\label{subsubsec:Ple_Ori_color_difference}

Differences in color-color relations between Orion and Pleiades, perhaps caused by differences in gravity or spots, may lead to biases in the fitting result. In this subsection, we first give an example on how the difference in color-color relation influences the fitting result.  We then discuss the effect of surface gravity and spots separately.

Figure \ref{fig:quantify_RP_J} shows $(BP-RP)$-$(RP-J)$ for Pleiades, Orion, and members of the beta Pic Moving Group, $\epsilon$ Cha association, $\eta$ Cham association, and TW Hya Association from \citet{Gagne2018,Gagne2018b}. To quantify $(RP-J)-(RP-J)_{\rm Pleiades}$ at a given $BP-RP$, $(RP-J)-(RP-J)_{\rm Pleiades}$ is given by setting several bins in $BP-RP$ and calculating the average $RP-J$ in each bin that includes a star. 

The $RP-J$ versus $BP-RP$ colors of the groups differ at $BP-RP\sim2$ ($\sim4000$ K or 3.6 dex), which may explain the temperature deviation in Figure \ref{fig:fitting_result_compare}.  As a simple example, only $BP$, $RP$, and $J$ pass-bands are used in the following demonstration. First, a star in Orion with $A_V \sim 0$ mag, $BP-RP=2.15$ ($\sim4000$ K or 3.6 dex) and $RP-J=1.62$ is selected. According to the Pleiades color-color relation and extinction vector in Figure \ref{fig:quantify_RP_J}, the fitted $BP-RP$ of this star is 1.32 ($\sim4800$ K or 3.68 dex). While the fitted temperature is higher, the fitted extinction is also higher, leading to a gap in $T_{\rm eff}$-$A_V$ relation at $\sim4000$ K. Although this is a simple demonstration using only three pass-bands, redder Orion stars may be one of the reasons for the temperature bump in Figure \ref{fig:fitting_result_compare}.

\begin{figure}[!t]
    \centering
    \includegraphics[width=0.9\columnwidth]{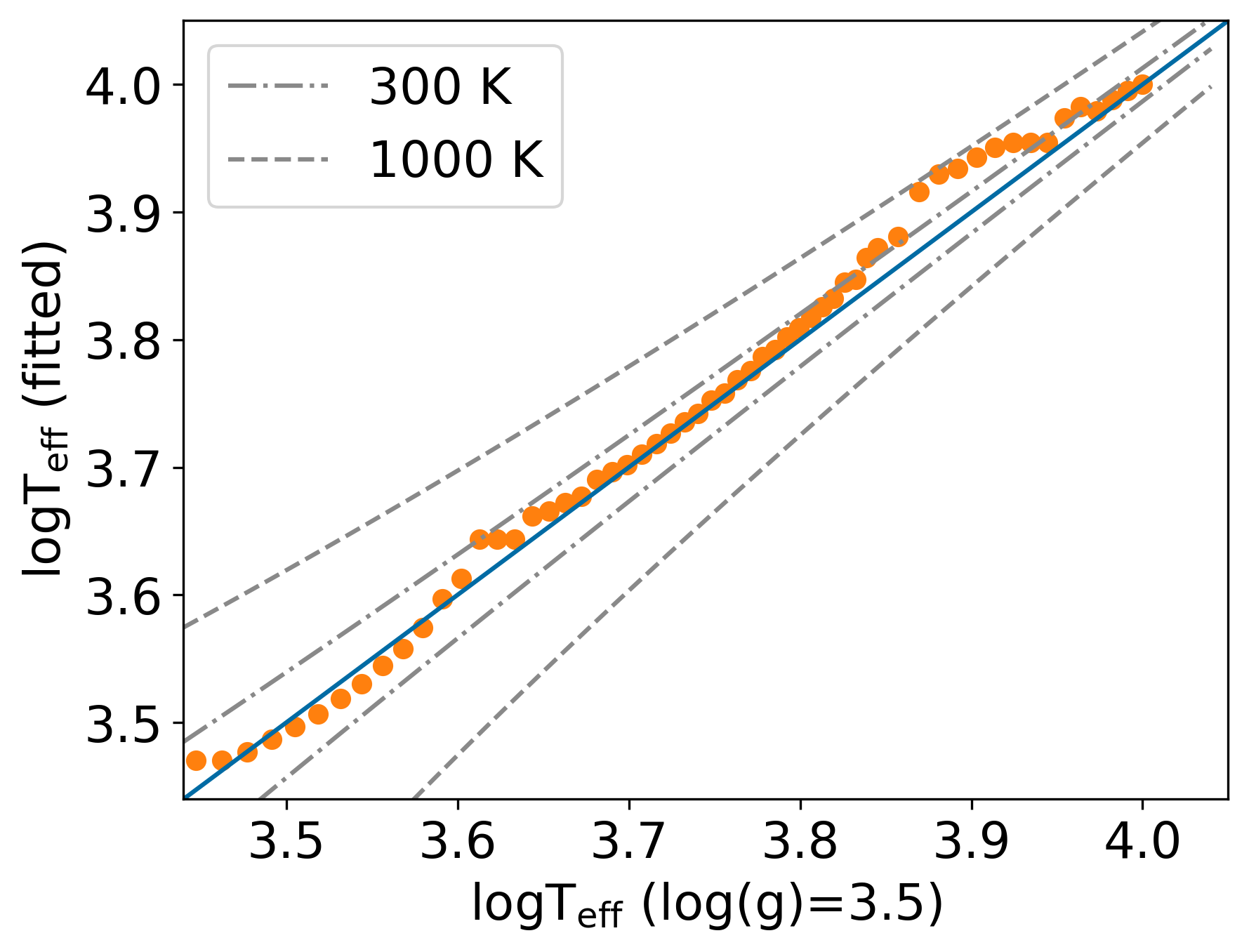}
    \caption{Comparison between temperatures for spectra with $\log g$ {=3.5} and temperatures from fits to those spectra using $\log g$ {=4.5} spectra as templates.}
    \label{fig:comparison_different_logg}
\end{figure}

These colors may be caused by the increase in surface gravity as a star contracts, with empirical measurements from APOGEE of $\log(g)\sim4.0$  for 2-6 Myr IC 348 stars  to $\sim4.8$ for $\sim$110 Myr Pleiades stars \citep{Cottaar2014}. To simulate the effect of surface gravity, we use BT-Settl models to describe that the lower gravity of younger stars leads to a higher fitted temperature at $\sim4000$ K.

BT-Settl spectra with $\log g$ {=4.5} are used to build basic SEDs as templates, while BT-Settl spectra with $\log g$ {=3.5} are then used as the targets to be fitted. {Offsets in Figure \ref{fig:comparison_different_logg} (lower $T_{\rm fit}$ at $\sim$3.55 dex and higher $T_{\rm fit}$ at $\sim$3.61 dex is consistent with the similar offset in Figure \ref{fig:fitting_result_compare}}. The offset at $3.6<\log {T_{\rm{eff}}}<3.7$ in Figure \ref{fig:fitting_result_compare} is at least partially attributed to the difference in surface gravity.

\begin{figure}[!t]
    \centering
    \includegraphics[width=0.9\columnwidth]{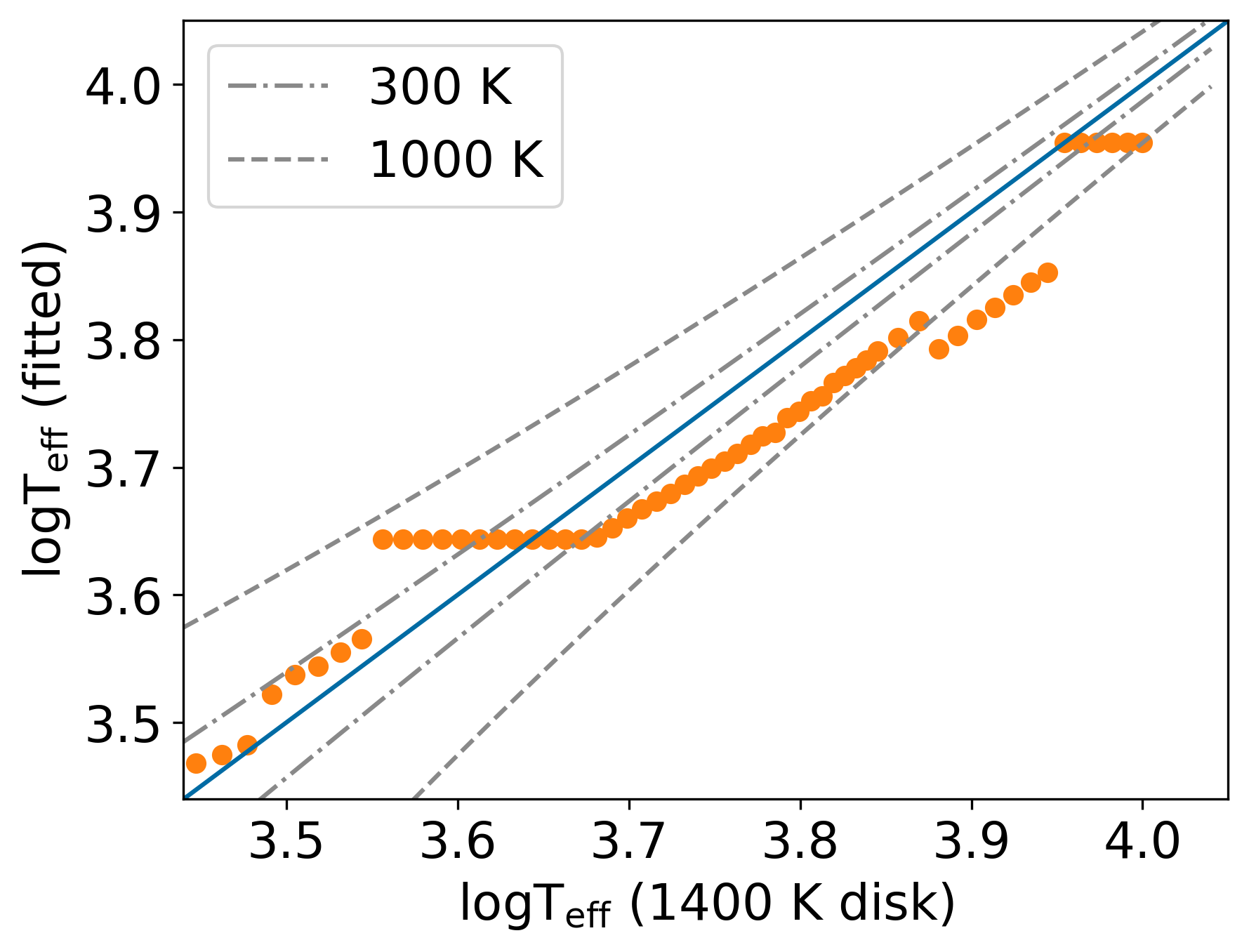}
    \caption{Comparison between the temperature of a BT-Settl spectra versus the fitted temperature of the spectrum plus a warm dust dusk.}
        \label{fig:comparison_disk_effect}
\end{figure}

In addition to surface gravity, spot properties also vary among YSOs \citep[e.g.,][]{Grankin2008} and influence SED fitting result. Spots on magnetically active cool stars redden the broadband SED \citep[e.g.][]{somers20}, leading to errors when fitting with single temperatures \citep{gully17}. Although the Pleiades stars are heavily spotted \citep[e.g.][]{stauffer03,Fang2016,Guo2018}, our SED fits may be insufficient in incorporating spots, if they are more important for young stellar objects \citep{Grankin2008}.

\subsubsection{{Fits to disk stars}}
\label{subsubsec:fits_to_disk_stars}

Stars with disks are not well described through photometric fits. Removing a few photometric points may avoid the influence of infrared excess and/or bluer optical colors caused by a disk, but enough pass-bands are still needed to constrain the photospheric properties.  Most fitting routines, including ours, do not incorporate accretion, which alters colors and in some cases dominates the { optical} emission \citep[e.g.][]{gahm08}. In such cases, any attempt to measure stellar properties from SEDs or spectra is dubious. More typical accreting stars, where the accretion flow affects but does not dominate the optical emission, have SEDs that are roughly simulated by models but are expected to show excess blue emission.

We evaluate uncertainties caused by emission from dust by creating SEDs from a combination of BT-Settl spectrum of the photosphere and a 1400 K blackbody for the dust, chosen as the dust sublimation temperature. The blackbody emission flux is scaled so that the $K$-band flux from the blackbody is {two times} of the photosphere flux \citep[consistent with the ratios in][]{Fischer2011}. The spectrum is then fit with the BT-Settl spectra alone, excluding $W1$ and $W2$. The SED templates are based on interpolating BT-Settl color-color relations. The temperature comparison (Figure \ref{fig:comparison_disk_effect}) shows that an offset of {$\sim1000 K$} is present for stars at $\log T_{\rm eff}\sim3.55$, while for the cooler stars the offset is smaller. 

Although this simple approach indicates that the fits are robust to the presence of steady dust disks, the effect of disks are expected to be more significant than described here. In some cases the disk will obscure the star. Accretion from the disk onto the star contributes and in some cases dominates optical colors. Both disk obscuration and accretion are variable.  These phenomenon are significant confounding factors that are not explored in this paper.

\subsubsection{{Fits to the Pleiades and constrained extinctions}}
\label{subsubsec:constrained_extinction}

\begin{figure}[!t]
    \centering
    \includegraphics[width=0.9\columnwidth]{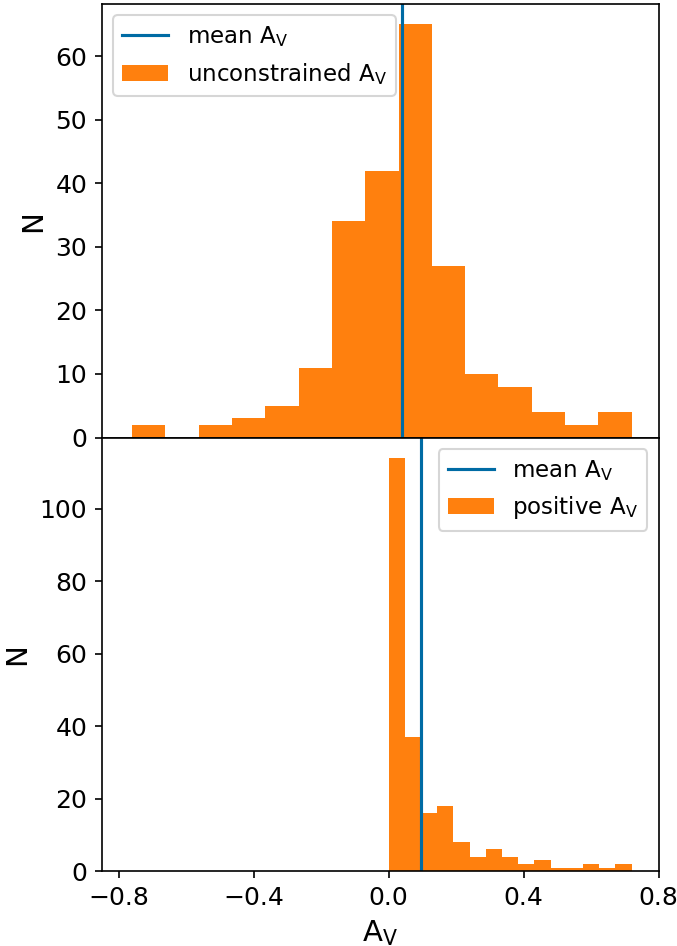}
   \caption{The distribution of best-fit $A_V$ (orange histogram; mean indicted with the blue line) of Pleiades stars, for fits that are unconstrained by reality and allowed to be negative (top panel) and fits that are constrained to be $\geq 0$ (bottom panel).}
    \label{fig:Pleiades_Av_distribution}
\end{figure}

The extinction range ($0 \leq A_V \leq 6$) affects the extinction distribution of a cluster with low extinction and statistical comparisons of low-extinction clusters to high-extinction clusters.  
Figure \ref{fig:Pleiades_Av_distribution} shows the best-fit $A_V$ to stars in the Pleiades.  The average $A_V$ is 0.04 mag when $A_V$ is allowed to be negative and 0.1 mag when $A_V$ cannot be negative.   The spread in estimated $A_V$ and therefore the size of this effect is likely larger for younger stars.

\subsubsection{Outliers in the temperature comparison}
\label{subsubsec:outliers_Teff_comparison}

In this subsection, we comment on stars with the best-fit SED temperature deviating by 1000 K from the previously measured literature temperature, with a total number of 13 stars after excluding those with photometric issues in Section \ref{subsec:method_test}:

For the four stars in this small sample with Pan-STARRS photometry, three are saturated in the $z$ band. Since $z$ band has the largest weight among the magnitudes when fitting SEDs (Section \ref{subsec:fit_photospheric_SED}, Table \ref{table:banduncertainties}), the saturated z-band photometry has a significant influence on the fitting result. The fourth star has a $g$ band image where the star appears to fall at the edge of the detector.

Six stars are variable stars reported by ASAS-SN \citep[the All-Sky Automated Survey for Supernovae,][]{ASASSN2014,ASASSN2017}, with amplitudes from 0.1 mag to 0.3 mag. A seventh star has not been reported as a variable star but shows a wavy light curve with an amplitude of $\sim0.1$ mag.

The previous two sets of stars have some overlap. Of the remaining stars, three with the highest $T_{\rm fit}$ have $A_V(\rm fit)<0.5$, indicating a higher degeneracy between $T_{\rm eff}$ and $A_V$, which is due to the extinction vectors becoming more parallel to color-color relations at higher temperature, as shown in the bottom panel of Figure \ref{fig:quantify_RP_J}.  Two stars are not explained by the above descriptions.

\begin{deluxetable*}{cccccccccccccccccc}
\label{table:Serpensfits}
\tablecaption{Fits of Serpens members}
\tablehead{\colhead{\textit{Gaia} {EDR3} ID} & \colhead{RA/deg} & \colhead{DEC/deg} & \colhead{Region} & \colhead{d/pc} & \colhead{T$_{\rm eff}$/K} & \colhead{$A_V$} & \colhead{$L$(log$L/L_{\odot}$)} & \colhead{Age/Myr} & \colhead{Mass/$M_{\odot}$} & \colhead{Disk} & \colhead{{Flag}}}
\startdata
4309522600884785536 & 289.69099 & 10.80327 & LDN 673 & 376 & 7014 & 0.84 & 0.69 & 21.8 & 1.50 & - & 1 \\
4309648735482263936 & 290.35284 & 11.11158 & LDN 673 & 414 & 4401 & 2.58 & 0.15 & 1.9 & 1.02 & 0 & 0 \\
4309847540917708416 & 290.04547 & 11.3505 & LDN 673 & 409 & 3013 & 2.03 & -0.51 & 1.0 & 0.18 & 0 & 0 \\
4309847884515108480 & 290.01491 & 11.37616 & LDN 673 & 406 & 3743 & 2.03 & -0.67 & 6.0 & 0.56 & - & 1 \\
4271424759186100352 & 275.28395 & -1.30382 & Distrib & 430 & 4282 & 1.90 & -0.10 & 2.6 & 0.86 & 0 & 1 \\
4271223720358239744 & 275.73095 & -1.40617 & Distrib & 369 & 4322 & 0.28 & -0.91 & 47.8 & 0.68 & 0 & 1 \\
4271528250719881728 & 275.50849 & -1.08444 & Distrib & 443 & 3005 & 1.29 & -1.35 & 3.8 & 0.09 & 0 & 0 \\
4269964371522804224 & 275.3711 & -2.62225 & Distrib & 449 & 4526 & 0.00 & -0.50 & 21.8 & 0.88 & 0 & 1 \\
4269965196156468480 & 275.51853 & -2.60617 & Distrib & 456 & 4215 & 0.12 & -0.66 & 22.9 & 0.82 & 0 & 1 \\
4270725160552584960 & 275.46313 & -2.40182 & Distrib & 446 & 4157 & 0.80 & -0.98 & 47.8 & 0.66 & 0 & 1 \\
\hline
\multicolumn{8}{l}{All photometry in a table for vizier}\\
\enddata
\tablecomments{"Flag": "0" represents that this star is not included in calculating cluster ages (Section \ref{subsec:Serpens_ages}).  Individual distances are adopted from \textit{Gaia} EDR3 parallaxes.}
\end{deluxetable*}

{\catcode`\&=11
\gdef\2016AandA...593A..99F{\citet{Feiden2016}}}

\section{SED Fits to Members of the Serpens Star-forming regions}
\label{sec:Serpens}
The primary motivation for this paper is to characterize the optically bright stars in Serpens. Of the previously identified optical members in Serpens,  569 stars pass the photometric selection criteria and have SEDs fit here. To analyze their SEDs, the first step is identifying if a star has a disk (Section \ref{subsec:Serpens_disk_star}). The SEDs are fit to the relevant bandpasses, with results and analysis presented in Section \ref{subsec:Serpens_fit_result_analysis}. We then analyze the cluster ages in Section \ref{subsec:Serpens_ages}.

\subsection{Disk Presence}
\label{subsec:Serpens_disk_star}
The presence of disks around young stars is typically assessed from mid-IR photometry.  Since much of the Serpens cloud was imaged by \textit{Spitzer}, we first cross-match the Serpens sample with \textit{Spitzer} catalogs \citep{Evans2003,Gutermuth2008,Dunham2015} and find 123 stars with \textit{Spitzer} data. A disk is determined to be present in 52 stars based on an IR spectral index $\alpha>-1.6$, following \citet{Dunham2015}.  
For the Serpens stars without \textit{Spitzer} data, disk presence is identified by WISE and 2MASS color cuts (Figure \ref{fig:disk_selection_KW2_HW1}). In total, there are 359 diskless stars (71 from \textit{Spitzer}, 288 from 2MASS and WISE colors as seen in Figure \ref{fig:disk_selection_KW2_HW1}), 86 disk stars (52 from \textit{Spitzer}, 34 from WISE), and 231 stars without disk information due to the lack of both \textit{Spitzer} data and reliable WISE data. 

To confirm consistency in assessing disk presence or absence, we evaluate 82 members with both {\it Spitzer} and WISE photometry. Applying these color cuts to regions imaged by {\it Spitzer} show that the methods are mostly consistent.  The 10\% of stars that change classification between disk and diskless, all located near the borders of the classification system.  To avoid the appearance of disk stars in fitting diskless stars, stars that have both WISE and {\it Spitzer} photometry are treated as diskless only if both sets of measurements indicate  the lack of disk.

Disk properties from Figure \ref{fig:disk_selection_KW2_HW1} are evaluated from the results of SED fits to Serpens members. The fitted $W1$ and $W2$ magnitudes are compared with observed magnitudes to evaluate the disk selection in Figure \ref{fig:disk_selection_KW2_HW1}. {We would then reclassify the presence of a disk if the $W1$ or $W2$ magnitude is at least $0.78$ mag or $0.98$ mag (the average magnitude difference of disk stars) above the photospheric level, while disk stars would be re-classified as diskless if the $W1$ and $W2$ magnitude are less than $0.05$ mag and $0.09$ mag (the average magnitude difference of diskless stars) of the photospheric level. In this reassessment, only one star is reclassified into the diskless category, and all stars previously identified as diskless remain diskless. We update results following this possible reclassification.}

\subsection{Isochrone fits to stellar properties}
\label{subsec:Serpens_fit_result_analysis}

\begin{figure*}[ht]
    \centering
    \includegraphics[width=0.9\textwidth]{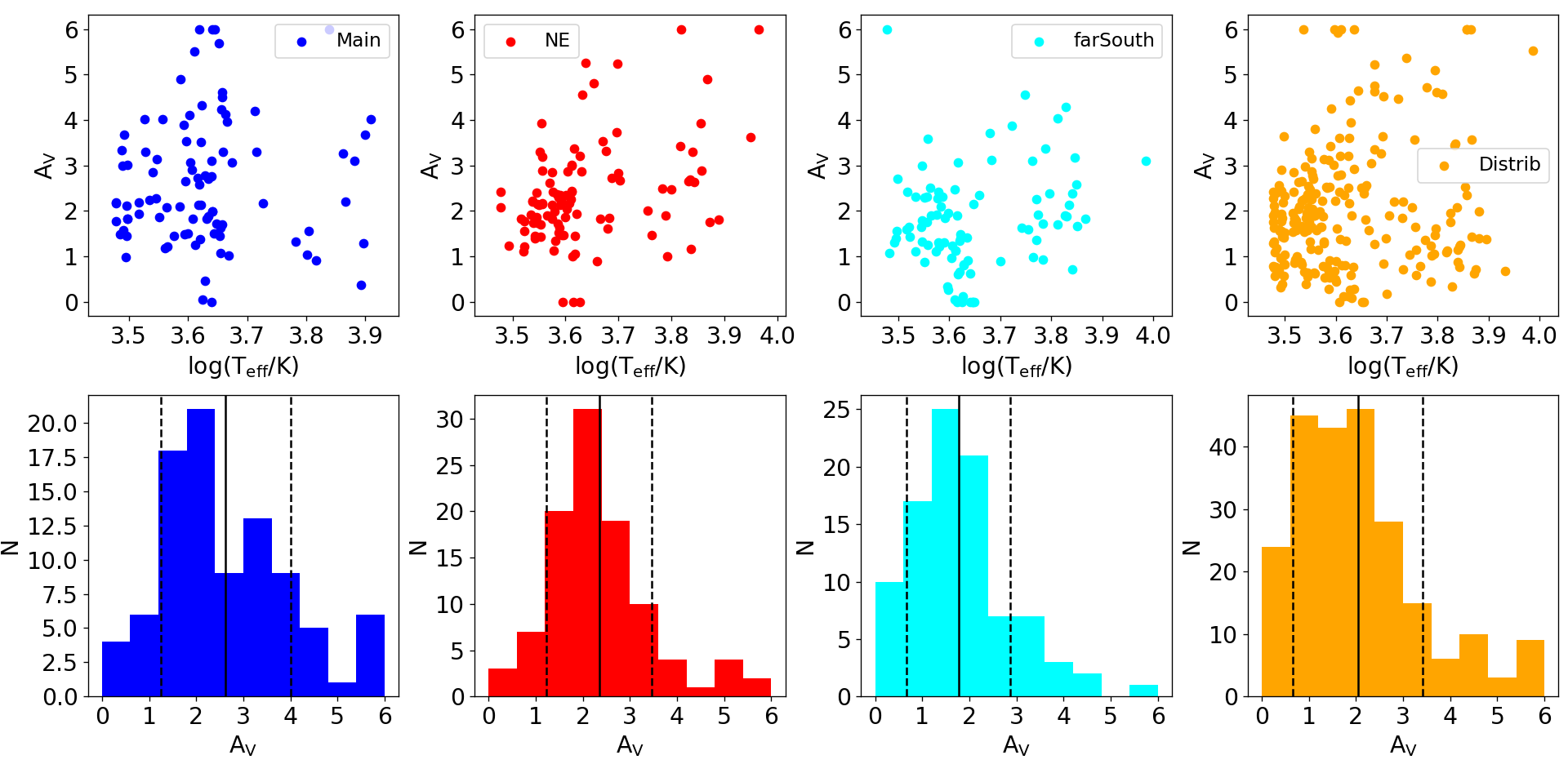}
    \caption{Temperature versus extinction (top panels) and the distribution of extinctions (bottom panels) for four major groups in Serpens. The black lines in bottom panels are the mean and standard deviation of $A_V$.}
    \label{fig:T_Av_Serpens}
\end{figure*}

\begin{figure*}[ht]
    \centering
    \includegraphics[width=0.9\textwidth]{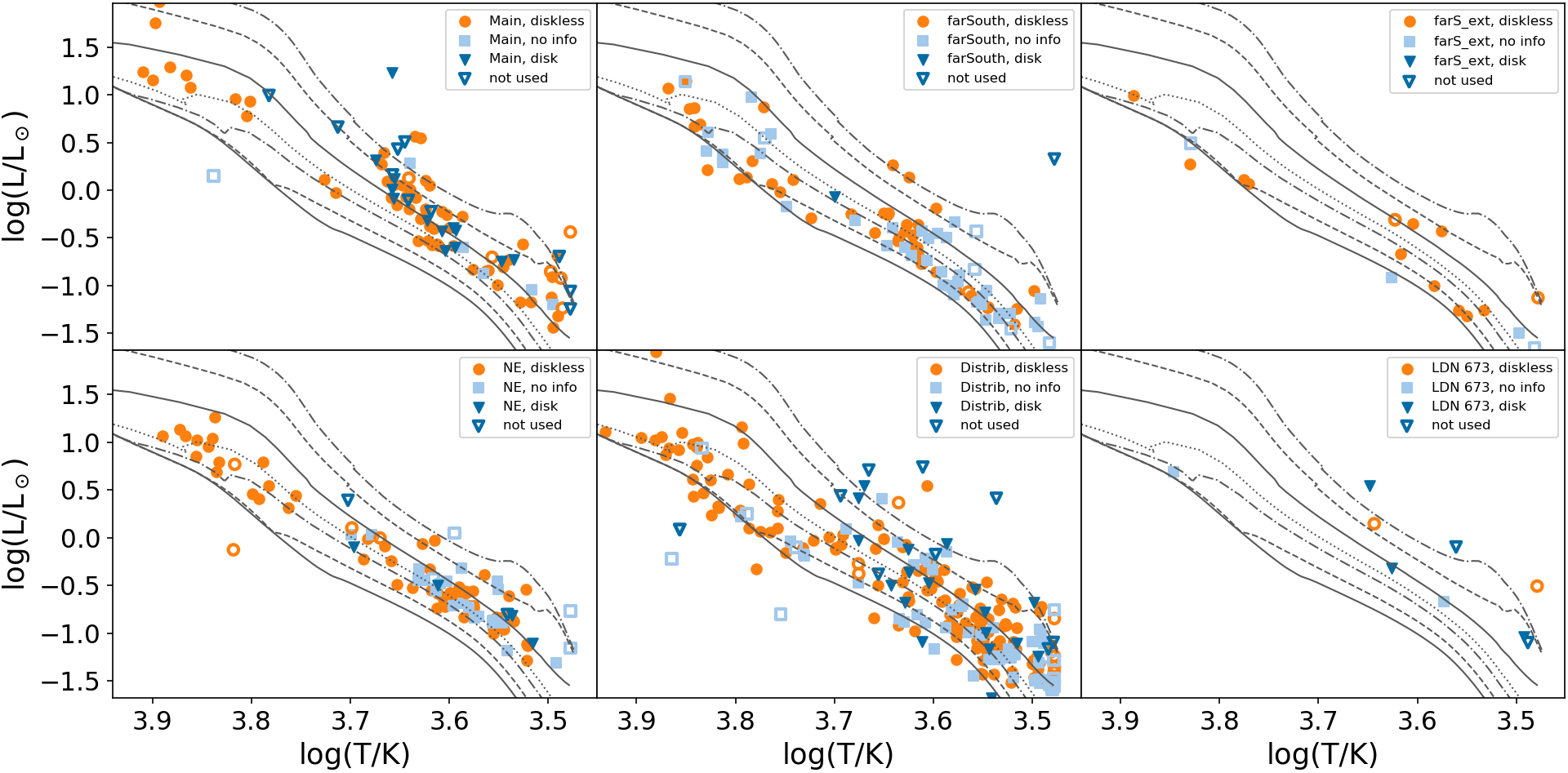}
    \caption{The HR diagram of Serpens members in different groups, compared with isochrones from the non-magnetic evolutionary tracks of   \citet{Feiden2016}, for stars with disks (deep blue triangles), without disks (orange dots), and without disk information (light blue squares). Unfilled symbols are stars which are not used in calculating ages. The gray curves are isochrones with 1 Myr, 2 Myr, 5 Myr, 10 Myr, 15 Myr, 30 Myr, 50 Myr from top right to bottom left.}
    \label{fig:HRD_Serpens}
\end{figure*}

\begin{figure}[ht]
    \centering
    \includegraphics[width=\columnwidth]{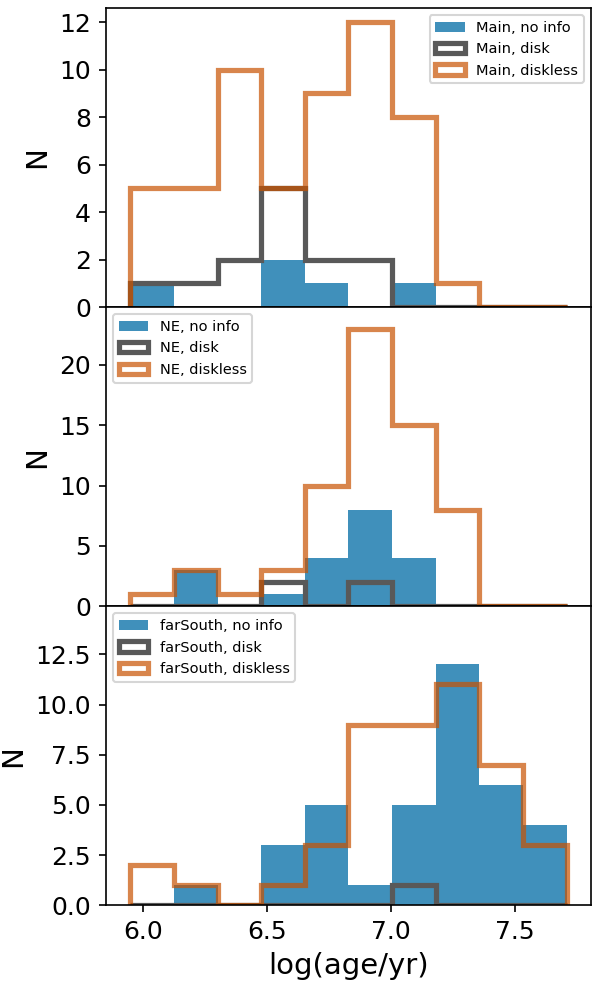}
    \caption{The distribution of ages for three groups in Serpens: Main (top panel), NE (medium panel) and far-South (bottom panel), in order of average age for the group.  The age spread in each group may not be significant \citep[see discussion in][]{soderblom14}.}
    \label{fig:age_num_dist}
\end{figure}

\begin{figure}[ht]
    \centering
    \includegraphics[width=\columnwidth]{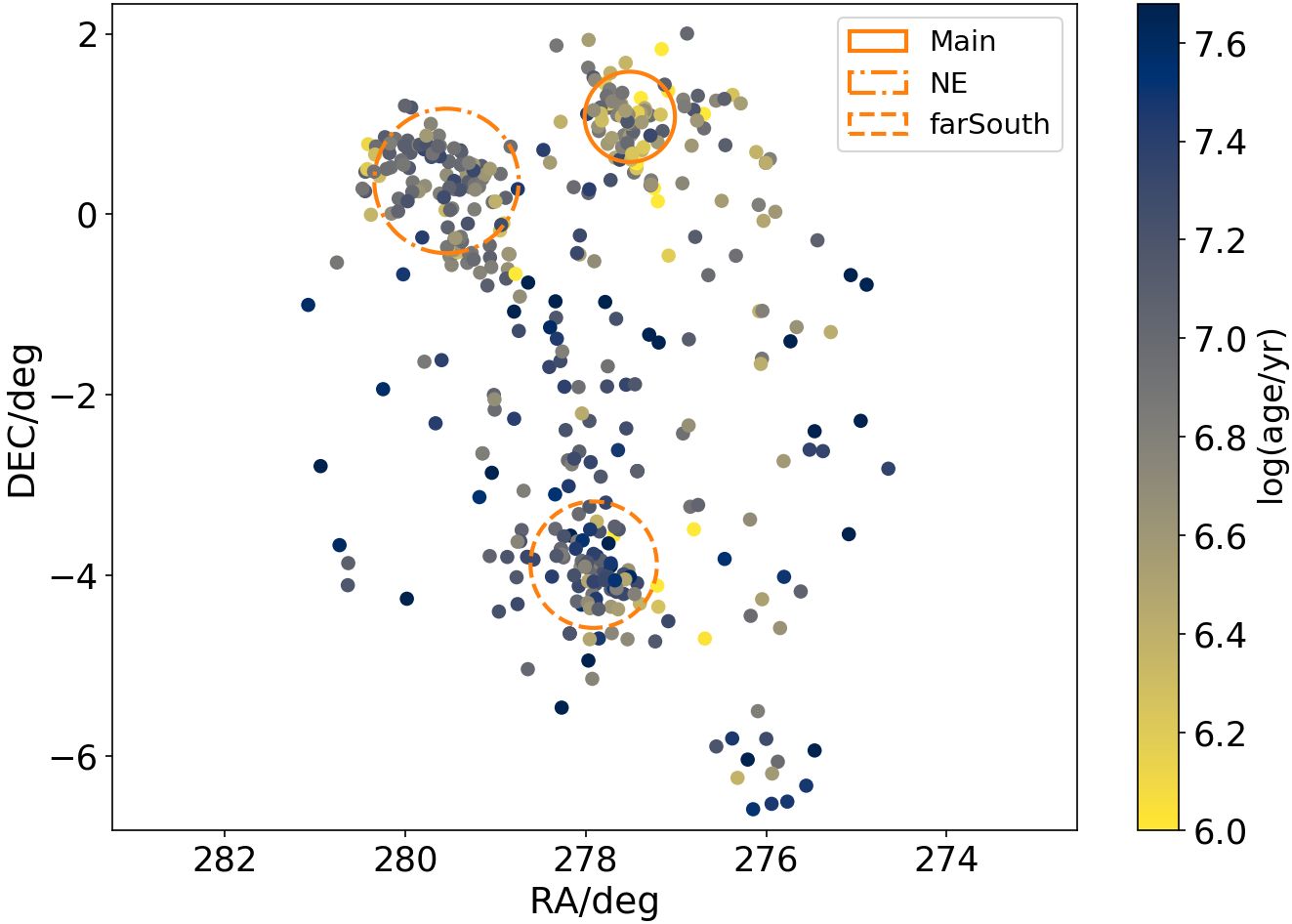}
    \caption{Spatial location of Serpens members with colors mapped by ages. Orange circles represent the location of three subgroups: Serpens Main, NE and farSouth, with location and radius adopted from Table 1 of \citet{Herczeg2019}.}
    \label{fig:age_map}
\end{figure}

Table \ref{table:Serpensfits} presents the best fit temperatures, luminosities, and extinctions for stars associated with the Serpens Molecular Cloud.   Figure \ref{fig:T_Av_Serpens} shows color excesses due to extinction versus temperature for stars in four different regions in Serpens (Main, NE, far-South, Distributed). The optical members of Serpens Main have the highest median extinction, while Serpens far-South is the least extincted, consistent with Serpens Main hosting more ongoing star formation than Serpens far-South.

The masses and ages of individual stars are determined by comparing the temperature and luminosity to pre-main sequence evolutionary models for non-magnetic stars calculated by \citet{Feiden2016}, using the isochrone fitting method in \citet{Zari2019}. Specifically, the likelihood for a single star to come from an isochrone with an age of $\tau$ is
\begin{equation}
    \label{eq:Likelihood_single_star}
    L(\tau) = \int L(\tau, m)dm,
\end{equation}
where $m$ is the mass of a point in the isochrone, and $L(\tau, m)$ is calculated from
\begin{equation}
    \label{eq:Likelihood_single_star_mass}
    L(\tau, m) = \frac{1}{(2\pi)^{1/2}\sigma_T} \frac{1}{(2\pi)^{1/2}\sigma_L} \exp{(-\chi^2/2)},
\end{equation}
with:
\begin{equation}
    \label{eq:chi_square_single_star}
    \chi^2 = \left(\frac{T-T_{iso}}{\sigma_T}\right)^2 + \left(\frac{Lum-Lum_{iso}}{\sigma_{Lum}}\right)^2,
\end{equation}
where $T$, $Lum$, $\sigma_T$, $\sigma_L$ represent temperature, luminosity and their errors (in log scale) of a star, while $T_{iso}$ and $Lum_{iso}$ are temperature and luminosity of a point in an isochrone which are determined by mass and age. The maximum $L(\tau)$ corresponds to the best-fit age. The isochrone fitting procedure is based on linear space of the isochrones.

\subsection{Age estimates for Serpens subclusters}
\label{subsec:Serpens_ages}

In this section, ages are estimated via different isochrones and different methods. {Stars with photometric problems or with best-fit parameters at the edge of the parameter space of our grid (Section \ref{subsec:method_test}) are excluded in calculating ages, with a total number of 455 stars used in age estimation.} The ages adopted in this study is derived in section \ref{subsubsec:adopted_ages}, while the comparison of ages are presented in Section \ref{subsubsec:age_other_iso}

\subsubsection{Ages adopted in this study}
\label{subsubsec:adopted_ages}

\begin{deluxetable*}{lccccccc}
\label{table:Serpens_age_iso_fit_std}
\tablecaption{Ages and errors (Myr) of Serpens groups with non-magnetic isochrones}
\tablehead{\nocolhead{region} & \colhead{disk} & \colhead{error} & \colhead{diskless} & \colhead{error} &  \colhead{total} & \colhead{error} & \colhead{disk fraction}}
\startdata
Main      & 3.0  & 0.5  & 4.0 & 0.2 & 4.0 & 0.2 & 0.22  \\
NE        & 7.6 & 2.9 & 8.3  & 0.3 & 8.3 & 0.3 & 0.03 \\
farSouth  & 9.5 & 9.3 & 11.5 & 0.4 & 13.8 & 0.9 & 0.04 \\
farS\_ext & - & - & 21.9 & 1.6 &  21.9 & 1.4 & 0.00  \\
FGAql     & - & - & - & - & - & - & -  \\
LDN673    & 2.9 & 0.5 & - & - &  6.3 & 1.4 & -  \\
Distrib   & 5.5 & 0.6 & 10.0 & 0.3  & 10.0 & 0.4 & 0.12 \\
\enddata
\tablecomments{Estimated with isochrone fitting.  Total includes disk and diskless, and sources with no disk information.}
\end{deluxetable*}

\begin{deluxetable*}{ccccccccccccc}
\label{table:Serpens_age_compare_iso}
\tablecaption{Ages and errors (Myr) of Serpens groups with different isochrones and different estimation methods}
\tablehead{\colhead{ } & \multicolumn{2}{c}{{Feiden, standard}} & \multicolumn{2}{c}{Feiden, magnetic} & \multicolumn{2}{c}{Feiden, magnetic} & \multicolumn{2}{c}{PARSEC} &  \multicolumn{2}{c}{PARSEC} & \multicolumn{2}{c}{PARSEC}\\
\colhead{ } & \multicolumn{2}{c}{$T_{\rm eff}-L$}  & \multicolumn{2}{c}{$T_{\rm eff}-L$}  & \multicolumn{2}{c}{$T_{\rm eff}-L$}  &\multicolumn{2}{c}{$T_{\rm eff}-L$} &  \multicolumn{2}{c}{$T_{\rm eff}-L$} & \multicolumn{2}{c}{CMD}\\
\colhead{ } & \multicolumn{2}{c}{Cool stars} & \multicolumn{2}{c}{All stars}   & \multicolumn{2}{c}{Cool stars} & \multicolumn{2}{c}{All stars} &  \multicolumn{2}{c}{Cool stars} & \multicolumn{2}{c}{Cool stars}\\
\colhead{Region} & \colhead{age} & \colhead{error} & \colhead{age} & \colhead{error} & \colhead{age} & \colhead{error} & \colhead{age} & \colhead{error} & \colhead{age} & \colhead{error} & \colhead{age} & \colhead{error}}
\startdata
Main      & 4.0 & 0.3 & 7.9 & 0.2 & 9.1 & 1.0 & 4.0 & 0.2 & 4.8 & 0.5 & 7.2 & 0.3 \\
NE        & 6.9 & 0.5 & 17.4 & 0.5 & 18.2 & 1.6 & 7.9 & 0.3 & 7.6 & 0.5 & 9.1 & 0.3  \\
farSouth  & 13.2 & 1.5 & 31.6 & 0.9 & 26.3 & 3.0 & 13.8 & 0.8 & 20.0 & 2.0 & 19.0 & 0.6 \\
farS\_ext & 9.5 & 3.3 & 34.7 & 2.0 & 20.0 & 6.2 & 22.9 & 2.3 & 15.1 & 4.5 & 25.1 & 1.9 \\
FGAql     & - & - & - & - & - & - & - & - & - & - & - & -   \\
LDN673    & 3.8 & 2.1 & 15.1 & 0.7 & 11.5 & 5.3 & 6.3 & 1.3 & 6.6 & 2.0 & 4.8 & 0.3 \\
Distrib   & 6.6 & 0.5 & 22.9 & 0.5 & 15.8 & 1.0 & 10.0 & 0.3 & 10.0 & 0.6 & 10.5 & 0.3  \\ 
\enddata
\tablecomments{Estimated with isochrone fitting.}
\tablecomments{$T_{\rm eff}-L$: Ages are estimated in temperature-luminosity space. CMD: Ages are estimated in color-magnitude diagram (Pan-STARRS $r-i$ versus $r$).}
\tablecomments{Cool stars: Stars with $r-i>1.2$ ($\sim3500< T_{\rm eff} <\sim4500K$).}
\end{deluxetable*}

The ages adopted in this paper are determined by comparing \citet{Feiden2016} non-magnetic isochrones to the temperature and luminosity of each source, adopting individual stellar parallax distances for each star (see Figure~\ref{fig:HRD_Serpens}.
Table~\ref{table:Serpens_age_iso_fit_std} lists the ages and errors of different groups within the cloud, as calculated from isochrone fitting. For the major groups in the cloud, Serpens Main is assessed an age of {4 Myr}, Serpens NE is {8.3 Myr}, Serpens far South at {13.8 Myr}, and the extended group  of far-South (labeled as farS\_ext) is 21.9 Myr. The small LDN 673 Group age of 6.3 Myr is calculated from 5 stars and is therefore not well constrained. 
Serpens South and W40 are deeply embedded and lack sufficient optical members for age estimates, but they are presumably younger than Serpens Main.

Figure \ref{fig:age_num_dist} shows the age distribution of Serpens Main, Northeast and far-South. To check if age distributions within a single subcluster are significant (see Figure~\ref{fig:age_num_dist}), age errors of individual stars are calculated via the method in \citet{Zari2019} {by adopting the 16th and the 84th percentiles of the likelihood (Equation~\ref{eq:Likelihood_single_star}) distribution. The average age error is $\sim 0.2$ dex. Since the error relies on temperature and luminosity errors from SED fitting, which are limited by fitting ranges (Section \ref{subsec:method_test}), the $\sim 0.2$ dex age error does not reflect the real age probability distribution}. Although some age spread within clusters is likely, it is unclear from our current methods and analysis whether we could subdivide any group according to age.
Adopting single distances for each cluster does not significantly change the age but would change individual stellar ages.

Figure \ref{fig:age_map} shows a map of members of the Serpens star-forming region, with stars colored by age.  Age difference among subgroups (Figure \ref{fig:age_num_dist} and Table \ref{table:Serpens_age_iso_fit_std}) is also indicated by Figure \ref{fig:age_map}. The youngest subgroup, Serpens Main, is the smallest and densest group, as shown in Figure \ref{fig:age_map}, while the oldest stars are scattered in the distributed region.

\subsubsection{Ages from other estimation methods}
\label{subsubsec:age_other_iso}

\begin{figure}[ht]
    \centering
    \vspace{5mm}
    \includegraphics[width=0.95\columnwidth]{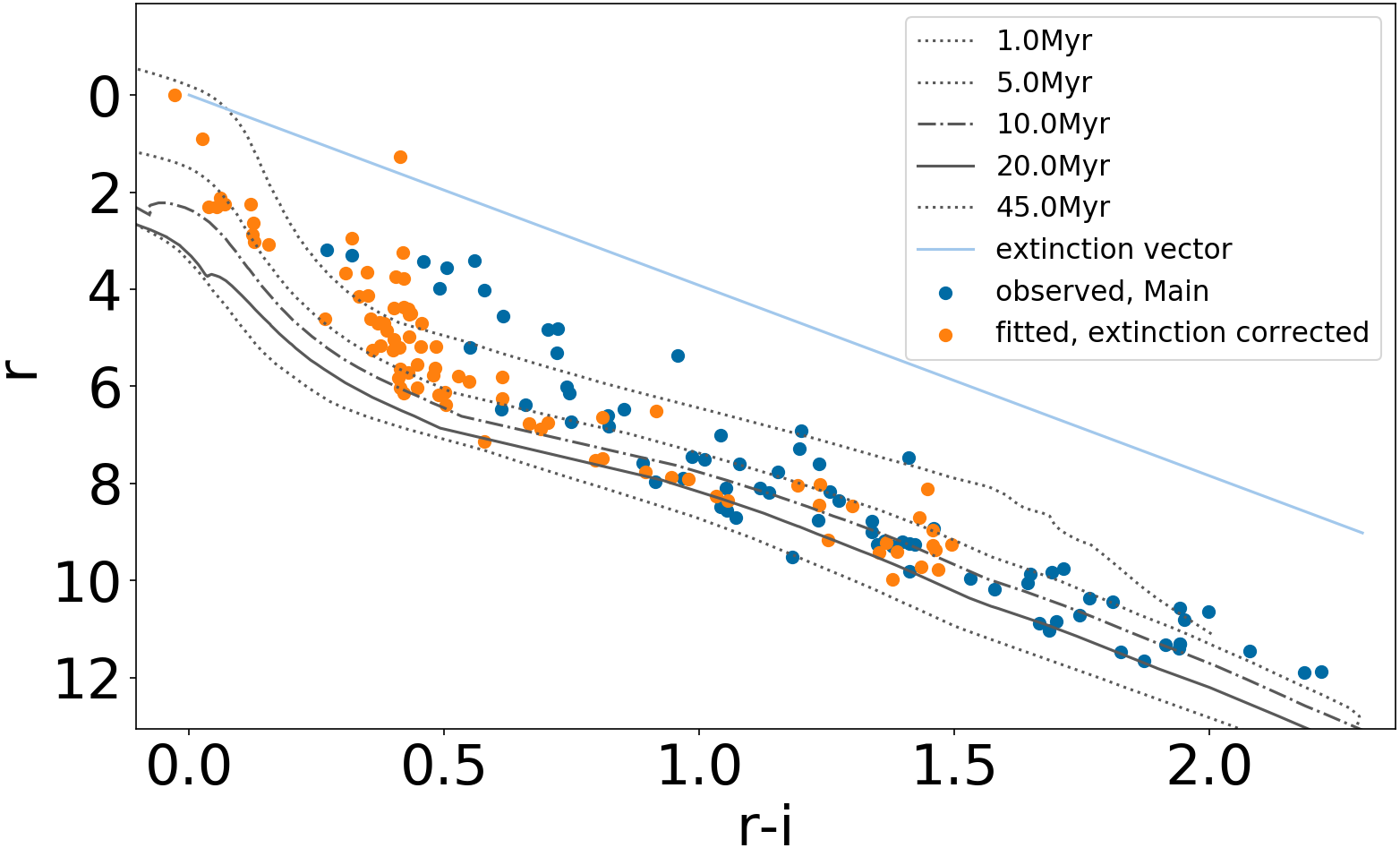}
    \caption{Color-magnitude diagram for Pan-STARRS $r$ versus $r-i$, with PARSEC isochrones (gray curves), with stars in Serpens Main (blue and orange dots represent the same set of stars with observed and extinction corrected result separately), and the extinction vector (light blue straight line) that runs parallel to the isochrones for cool stars.}
    \label{fig:CMD_PS1_Main}
\end{figure}

\begin{figure*}[ht]
    \centering
    \includegraphics[width=\textwidth]{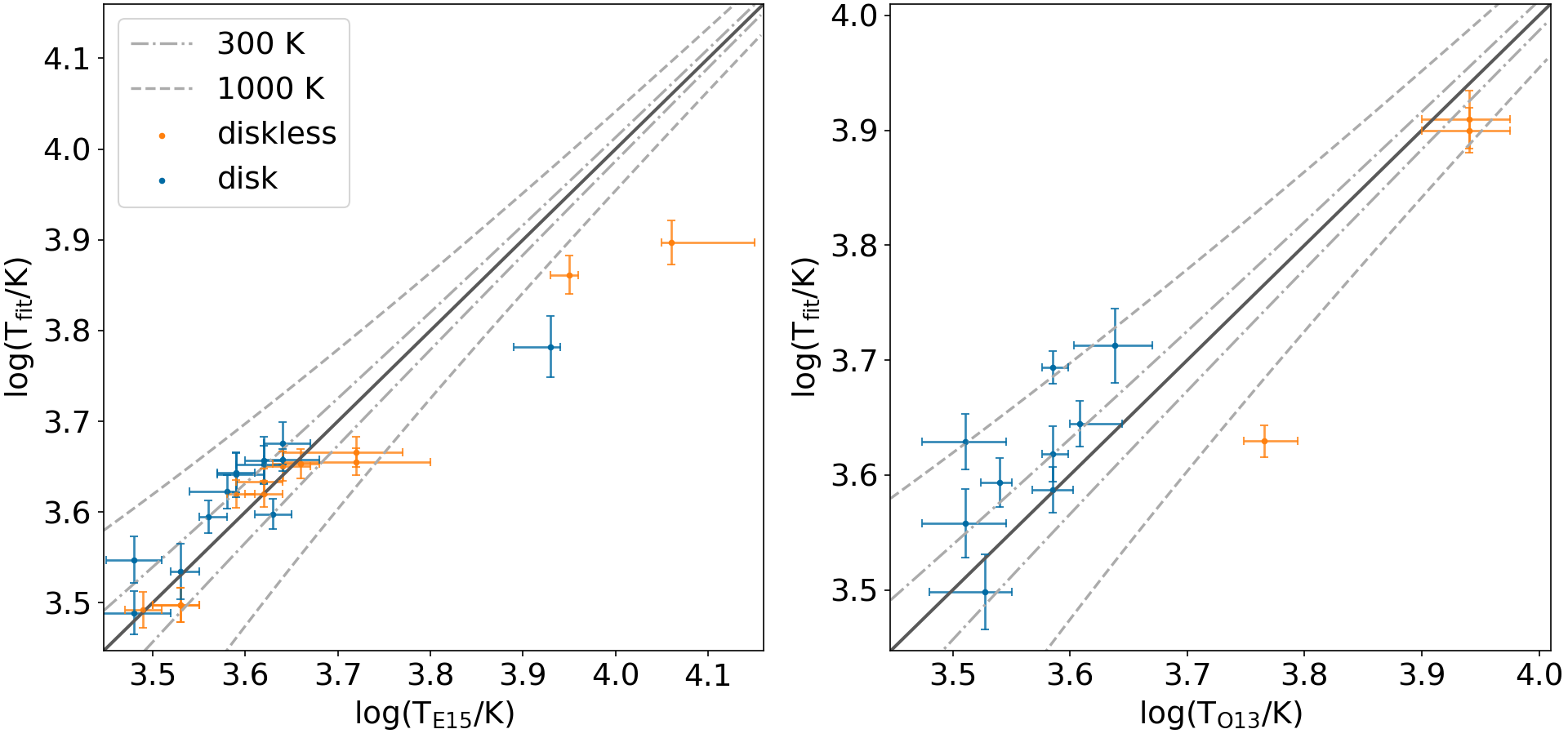}
    \caption{Comparison between temperatures derived in this paper and in \citet{Erickson2015} (left) and \citet{Oliveira2013} (right). Orange symbols are diskless stars, and blue symbols are disk stars.}
    \label{fig:E15_T_compare}
\end{figure*}

The age estimates are sensitive to evolutionary tracks and to the method for isochronal fitting.  In this section, we explore systematic uncertainties in the ages due to the methodological choices. 

Table \ref{table:Serpens_age_compare_iso} compares 
 ages from the non-magnetic and magnetic \citet{Feiden2016} and from PARSEC tracks \citep{Bressan2012, ChenYang2014, ChenYang2015, TangJing2014}.   In general, the assessed ages {of the magnetic \citet{Feiden2016} evolutionary models} are older {than ages from non-magnetic models}, due in part to the larger initial stellar radius in the magnetic models. The PARSEC ages are mostly consistent with the \citet{Feiden2016} non-magnetic ages.

Studies are often limited to age estimates from color-magnitude diagrams and may lack extinction values that are needed for the placement of stars on HR diagrams.  Some color-magnitude diagrams (including $r-i$ versus $r$ (Figure \ref{fig:CMD_PS1_Main}) have $r$ versus $r-i$ isochrones that are parallel to the extinction vector over some color (temperature) range, allowing for age estimates with limited information \citep[e.g.][]{sicilia04}. The isochronal fits to these color-magnitude diagrams lead to ages that are older than HR diagram fits (Table \ref{table:Serpens_age_compare_iso}, the last four columns), likely because the $r$ versus $r-i$ isochrones are not exactly parallel to the extinction vector. For example, a star with $r-i=1.3$ and $M_r=8.8$ is located along the 10 Myr isochrone, but if shifted for $A_V=2$ mag would then have $r-i=0.86$ and $M_r=7.1$, which corresponds to $\sim7$ Myr. Thus, ages from observed $r-i$ versus $r$ are slightly older than extinction corrected ages.

\section{A Revised Star Formation History of the Serpens Molecular Cloud}
\label{sec:discussion}

In Section \ref{subsec:Serpens_ages}, we estimate the following ages from \citet{Feiden2016} non-magnetic isochrones for the following sub-clusters: Serpens Main at {4 Myr}, Serpens NE at {8.3 Myr}, Serpens far-South at {13.8 Myr}, Serpens farS\_ext at {21.9 Myr}, the distributed population at {10 Myr}, and LDN 673 Group at {6.3 Myr}.  The youngest regions, W40 and Serpens South, have few optical members identified for this study, so their ages are not assessed in this work.  Serpens Main also has a large protostar population that is younger than the optical population evaluated here.  In Appendix~\ref{appendix:coll359}, we apply similar techniques to the nearby open cluster Collinder 359 and assess an age of 21 Myr.  

In this discussion, we compare our ages with literature ages, place the optical sources in the context of more complete catalogs, and discuss these results in the context of star formation within the Serpens cloud complex.

\begin{deluxetable*}{ccccccccc}[t]
\label{table:Serpens_space_motion}
\tablecaption{Proper motion and radial velocity of Serpens groups and Collinder 359.}
\tablehead{\nocolhead{region} & \colhead{PM(R.A.)} & \colhead{$\sigma$ (PM R.A.)} & \colhead{PM(Dec.)} & \colhead{$\sigma$ (PM Dec.)} &  \colhead{RV} & \colhead{$\sigma$ (RV)} & \colhead{$N_{RV}$\tablenotemark{$1$} }\\
\nocolhead{region} & \colhead{mas/yr} & \colhead{mas/yr} & \colhead{mas/yr} & \colhead{mas/yr} &  \colhead{km/s} & \colhead{km/s} & \colhead{}}
\startdata
Main      & 3.09 & 0.40 & -8.48 & 0.46 & -9.6 & 2.4 & 2  \\
NE        & 2.58 & 0.33 & -8.13 & 0.36 & -12.1 & 5.5 & 4  \\
farSouth  & 1.94 & 0.36 & -8.90 & 0.31 & -2.4 & 2.1 & 5  \\
farS\_ext & 1.76 & 0.50 & -8.66 & 0.70 & - & - & -  \\
FGAql     & 1.67 & 0.19 & -9.74 & 0.28 & - & -  & - \\
LDN673    & 2.56 & 0.55 & -10.23 & 0.23 & - & - & - \\
Collinder 359   & 0.64 & 0.16 & -8.72 & 0.06 & 10.8 & 4.9 & 11 \\
\enddata
\tablenotetext{1}{Number of stars which are used to calculate the RV of a group.}
\end{deluxetable*}

\subsection{Comparisons of stellar parameters to results from other surveys}
\label{subsec:compare_other_Serpens}

Several Serpens stars in our study have photospheric properties assessed previously with spectroscopic data. This section briefly compares our results to those literature results.

\citet{Erickson2015} used optical imaging of Serpens Main to select candidate YSOs from locations in the color-magnitude diagram and then evaluated membership of 700 candidates with moderate resolution optical spectra. Of their {63} optical members, {26} have {\it Gaia} DR2 proper motions or parallaxes that are inconsistent with Serpens, nine were identified as non-members from {\it Gaia} DR2 astrometry by \citet{Herczeg2019}, two lack sufficient photometry for our fits (see Table~\ref{table:photometry_criteria}), {and one has no spectral type in \citet{Erickson2015}.}  Figure~\ref{fig:E15_T_compare} compares the temperatures from here and \citet{Erickson2015} for the { remaining} {25} stars in both studies.  {Three} stars are at the upper limit of $A_V$ ($\sim6$ mag) of the fitting, which is consistent with their literature $A_V$ ({5.5-6.5} mag in \citealt{Erickson2015}, 5-9 mag in \citealt{Getman2017}). For cool stars ($<5000$ K), {the fitted $T_{\rm eff}$ here generally agrees with \citet{Erickson2015} measurements, with an average difference of $\sim$250 K.} {Three} hot stars have temperatures that are {$>$2000 K} cooler here than in \citet{Erickson2015}.
 With \citet{Erickson2015} $T_{\rm eff}$ and luminosity,and \citet{Feiden2016} non-magnetic isochrones, the members in \citet{Erickson2015} have an age of 2.8 Myr. 
For these overlapping stars, the derived ages are 3.7 Myr (\citet{Erickson2015} $T_{\rm eff}$ and luminosity) and 3.9 Myr (fitted $T_{\rm eff}$ and luminosity), indicating consistency in results. The younger overall cluster age obtained by \citet{Erickson2015} is likely attributed to differences in selection.

\citet{Oliveira2013} measured spectral types and temperature of candidate Serpens YSOs with optical spectroscopy and used \textit{Spitzer} infrared photometry to study disk properties. Figure \ref{fig:E15_T_compare} compares the temperatures of the stars that overlap between our sample and \citet{Oliveira2013}. Cool stars ($<\sim5000$ K) have fitted $T_{\rm eff}$ {$\sim$500 K} hotter than in \citet{Oliveira2013} while hot stars have similar $T_{\rm eff}$. Using the \citet{Feiden2016} isochrones, the ages of stars in \citet{Oliveira2013} are 1.8 Myr for all members. For stars that overlap between our samples, the average age is 1.7 Myr with $T_{\rm eff}$ and $L$ from \citet{Oliveira2013} and 3.8 Myr for properties measured here, a significant discrepancy { that indicates methodological differences.}

\subsection{Comparison to the distribution of CO gas}
\citet{SuYang2020} mapped CO towards the Aquila Rift region, which { covers} the Serpens star-forming region in our study (Serpens NE, farSouth and LDN 673). 
Strong CO emission found by \citet{SuYang2020} is consistent with the relatively young ages of optical members in Serpens NE and LDN 673. In Serpens far-South (also labeled as Sh 2-62 in \citealt{SuYang2020}, and also known as MWC 297 region, \citealt{rumble15}), the strong CO emission region is coincident with the members that are bright in the infrared and faint or undetected at optical wavelengths (see also Figure 4 of \citealt{Herczeg2019}).   Although these regions are adjacent to each other, they are not necessarily related; the optical members that we associate with Serpens far-South are less extincted and spatially offset from the young stellar objects identified by \citet{Dunham2015}.
The older ({$\sim13.8$} Myr) optical members are not in the ongoing star-forming region of Serpens far-South and are not associated with molecular gas. The LSR velocities of $^{12}$CO in Serpens NE and far-South is between 8 km/s and 12 km/s (Figure 5 of \citealp{SuYang2020}), corresponding to $\sim-8$ km/s to $\sim-4$ km/s in heliocentric coordinate. {\textit{Gaia} radial velocities of the two groups (Table \ref{table:Serpens_space_motion}) are $<1\sigma$ away from the boundary of $^{12}$CO velocities.}

\begin{figure*}[ht]
    \centering
    \includegraphics[width=\textwidth]{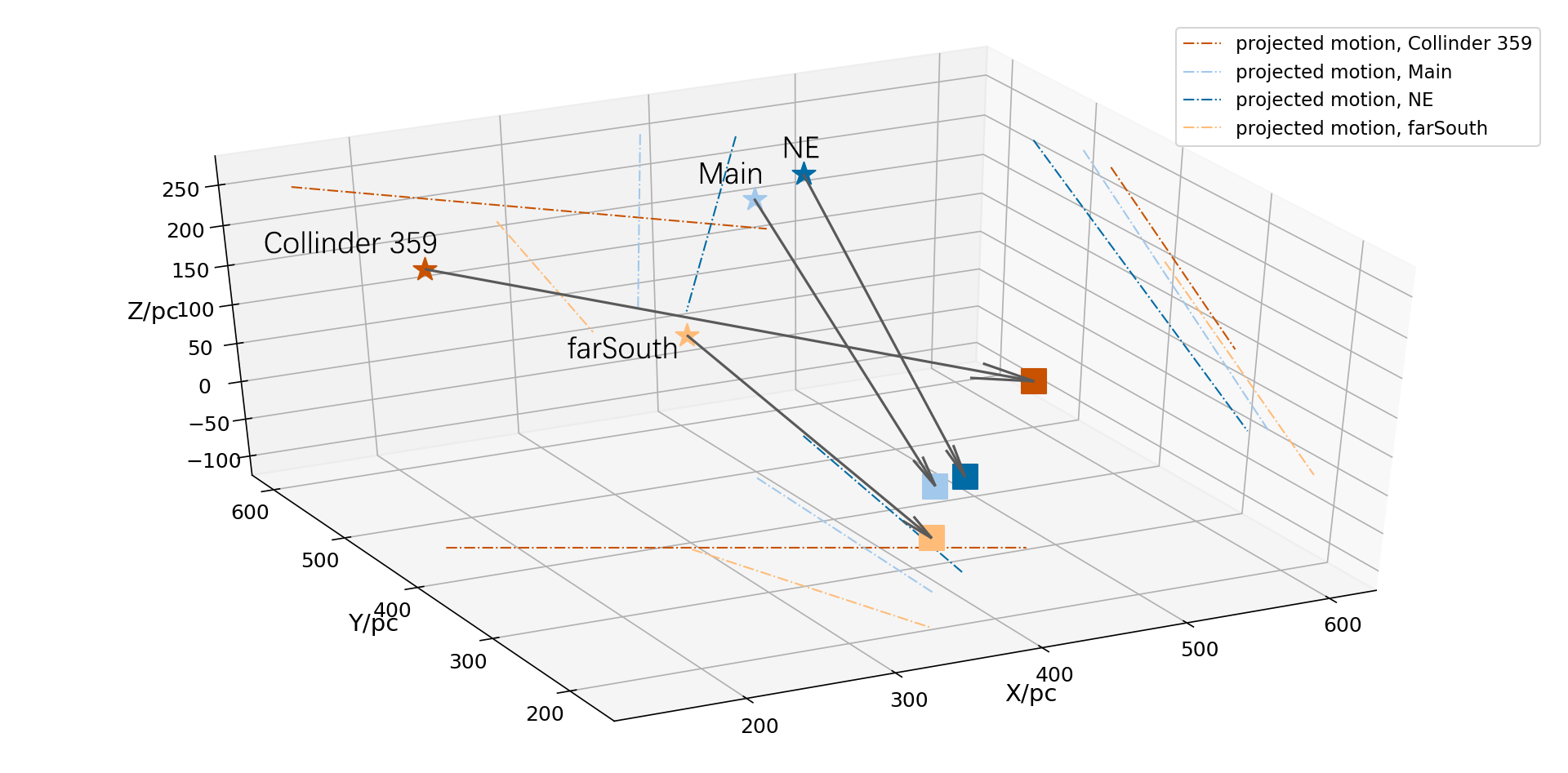}
    \caption{Current position (squares) and inferred position 15 Myr ago (star symbols) (Galactic coordinates, the Sun is at (X,Y,Z) = 0,0,0) of three groups in Serpens and nearby open cluster Collinder 359. Serpens Main is younger than 15 Myr but is still presented. Black arrows represent the space motion of the groups. Colored dashed lines are the projection of space motions into (X,Y), (X,Z) and (Y,Z) planes.}
    \label{fig:3D_plot}
\end{figure*}

\subsection{Co-location of disks and protostars}
\label{subsec:disk_fraction}

The disk fraction of a cluster provides an indication of cluster age, relative to the disk fraction in other clusters \citep[e.g.][]{Haisch2001,hernandez08,Bell2013}.  In Serpens, some optically bright members are in clusters that have protostars, especially in the active star-forming regions Serpens Main and Serpens NE.  Other Serpens members are far from any known protostars but are in groups that still have disks (see Figure 4 of \citealt{Herczeg2019}).

The overall disk fraction for the optical members of the Serpens star-forming region is 49 of 406 stars (12\%), from the sample of high confidence members with reliable photometry (samples without WISE photometry are excluded).  This disk fraction suggests ages of 5--10 Myr, roughly consistent with our age estimates.

This disk fraction applies only to the optical members and is not necessarily representative of the entire cluster.  An analysis based on the full membership requires significant assumptions and extrapolations.
For Serpens Main, the high-confidence members include 17 stars with disks and 62 diskless stars (22\% disk fraction, as listed in Table \ref{table:Serpens_age_iso_fit_std}).  The total statistical population consists of 265 members, as measured from the excess population associated with the cluster \citep{Herczeg2019}.  If we assume that the populations are similar, then the optical population would include 72.7 disk stars and 192.3 diskless stars.  From infrared surveys, \citet{Dunham2015} identified 222 likely members (excluding AGB stars).  Of the stars with {\it Gaia} astrometry, 50\% of the evolved sources and 86\% of the 73 disks are consistent with Serpens membership.  In follow-up with the envelope tracer HCO$^+$, \citet{heiderman15} confirmed that 88\% of candidate protostars in Serpens have an envelope, confirming membership and evolutionary stage.  If we assume that all matches are in the statistical population to avoid double-counting, then we find that Serpens Main consists of 45.5 protostars, 122.6 stars with disks, and 197.8 diskless stars.  This estimate includes significant uncertainties, including complications from binarity and extrapolations from the well-defined optical sample to the infrared sample, which may be younger.  The disk fractions may be underestimated if many diskless sources have high extinctions, either because they have high extinctions or because they are located within or behind the cloud, and were not selected in  the \citet{Dunham2015} catalog, 

Based on the above arguments, we adopt for Serpens Main a disk fraction of 38\% and a disk+protostar fraction of 46\%.  Similar estimates lead to disk+protostar fractions of 12\% for Serpens NE and 16\% for Serpens far-South.  These disk fractions support our relative ages: Serpens Main is younger than Serpens NE and Serpens far-South.  While these disk fractions indicate average ages for the clusters, the presence of protostars, disks, and diskless stars in the same cluster are likely the consequence of an age spread within the cluster \citep[see, e.g.][]{kristensen18}.

\subsection{Dynamics of the subclusters in and around the Serpens Molecular Cloud}
\label{subsec:subgroup_dynamic}

The identification of optically bright members of the Serpens molecular cloud complements previous selections from the infrared. The infrared selections identified several star clusters that are deeply embedded in molecular clouds \citep[e.g.,][]{Winston2009,Kuhn2013,Dunham2015}. The optically bright members are lightly extincted, some still in clusters and some distributed across the region. The infrared members trace the ongoing star formation, while the optical members trace star formation in the recent past. The ages of these different samples help to quantify when and where the bursts of star formation occurred. 

To infer the formation history of Serpens, ages are combined with space velocities of these Serpens groups (Table \ref{table:Serpens_space_motion}), with proper motions obtained from \citet{Herczeg2019} and the weighted mean radial velocities and associated uncertainty \footnote{The radial velocity uncertainty is the larger value between the weighted mean uncertainty and standard deviation.} obtained from members with velocities in \textit{Gaia} DR2. The radial velocities of Serpens Main, NE, farSouth are $-9.6 \pm 2.4$ km/s, $-12.1 \pm 5.5$ km/s, and $-2.4 \pm 2.1$ km/s separately. However, only few stars have radial velocity data (listed in Table \ref{table:Serpens_space_motion}), and the radial velocity uncertainties are significantly larger than proper motion uncertainties ($\sim 0.2$ mas, corresponding to $0.4$ km/s). Thus, the radial velocities are far less reliable than proper motions, and the radial velocity discrepancy between stars and gas is likely due to the uncertainties of stellar velocity.  The radial velocity of Collinder 359 is calculated from \citet{Cantat-Gaudin2018} samples with membership probability $P_{memb}\geq0.9$ and Gaia radial velocity, leading to $10.8 \pm 4.9$ km/s.\footnote{This radial velocity is consistent within uncertainties to the radial velocity of $8.30 \pm 1.79$ km/s calculated from the GALAH survey \citep{Carrera2019} and $5.28 \pm 3.25$ km/s from Gaia DR2 using sources in \citet{Cantat-Gaudin2018} with $P_{memb}\geq0.4$ \citep{Soubiran2018}. The \citet{Soubiran2018} radial velocity rejected samples deviating more than 10 km/s from the mean.}

With these space motions, the relative locations of the sub-clusters can be traced back in time (Figure \ref{fig:3D_plot}). The two youngest groups, Serpens Main and Serpens NE ({4} Myr and {8.3} Myr), have been co-moving from their birth to present, separated by a distance of {$\sim60$} pc (4 Myr ago) to $\sim40$ pc now, indicating that they may have similar formation history. The highly embedded regions W40 and Serpens South lack optical members in this study and are  expected to be in the same cloud complex as Serpens Main, as suggested by \citet{Ortiz-Leon2017}. The older group, Serpens far-South (17 Myr), has larger velocity difference from Serpens Main (e.g., 4.6 km/s in Galactic X coordinate, 3.2 km/s in Y, and 3.7 km/s in Z). Thus, Serpens far-South may have a different origin from Serpens Main ($\sim 160$ pc away from Serpens Main at the birth of far-South) or had star formation triggered by some interaction that had a different velocity at that location.
The open cluster Collinder 359 reached a minimum distance of $\sim60$ pc to Serpens Main at $\sim5$ Myr ago, when Serpens Main was born; whether such condition is coincidental or it reflects possible connections between Collinder 359 and Serpens is unclear.

The initial analysis of Gaia DR2 astrometry of the Serpens region by \citet{Herczeg2019} also includes a discussion of ages and kinematic properties of the sub-clusters. In Serpens NE, \citet{Herczeg2019} find that some stars are located outside of the molecular cloud, either because stars were born in a centralized space and were flung away (cluster dynamics) or because feedback from the older stars eroded nearby cloud and changed the birth place of younger stars. If these stars were born at the same spatial location, the velocity dispersion would require that they are $\sim 20$ Myr old.  With the age of Serpens NE of 9 Myr, that scenario is more plausible { than if the stars were much younger.}

In addition to Serpens, some other nearby young star-forming regions also contain several subgroups with different kinematic properties. For example, \citet{Krause2018} studied stars and gas of the Scorpius–Centaurus association and described the following sequential scenario to explain the formation history of the subgroups in this association: the first star-forming activity generated superbubbles, which triggered more star-forming events, with subgroups that move in different directions due to gas interaction. \citet{Krause2018} also suggested that the scenario might apply to many young star-forming regions in the Milky Way. If Serpens has a similar formation scenario, the first star-forming event would start at $\sim26$ Myr ago (Serpens farS\_ext and farSouth), which triggered the second star-forming activity at $\sim9$ Myr ago (Serpens NE and Main), and the current star-forming events are in W40 (the most embedded region), Serpens Main and NE.

 Multiple star-forming events in a single cluster appears to be the common mode of star formation.  In a \textit{Gaia} DR2 study of the Orion OB association , \citet{Zari2019} found in that multiple star-forming events occurred and leaded to the subgroups. In a search for stellar groups in the Taurus field, \citet{Liu2021} identified 8 groups younger than $\sim4$ Myr and 14 older groups (8-49 Myr).  In a smaller region, \citet{esplin22} found older stars in Corona Australis and inferred a past burst of star formation, likely related to the ongoing star formation in the Coronet Cluster.  These processes seem common to nearby low-mass star formation.

\section{Conclusions}
\label{sec:conclusion}

We develop an SED fitting method for diskless young stellar objects based on empirical color-color relationships from the Pleiades and temperatures measured from APOGEE near-IR spectra by \citet{Cottaar2014}. 
When using stars in Orion to test the method, the fitted temperature is consistent with temperatures from APOGEE spectra \citep{Kounkel2018} to within {$\sim$240 K} for diskless stars and {$\sim$630 K ($\sim$170 K) for disk stars (with $T_{fit}<$5000 K), all substantial improvements on SED fitting from other commonly used methods.}  
Shortcomings in the method include extinction coefficient variation, dust emission from stellar disk, different surface gravity, and different color-color relations among YSOs.

We collect photometry for high-confidence optical members of the Serpens Molecular Cloud and then fit SEDs with our empirical color-color relationships to estimate temperature and luminosity for each star.  The stars belong to several distinct groups in and around the cloud.  Subcluster ages range from {$\sim 4$ Myr}  for Serpens Main to {$\sim 22$ Myr} for dispersed stars located to the south of the cloud for standard non-magnetic evolutionary tracks. The optically bright population is much older than the populations that are revealed from infrared selection criteria, such as the Serpens South region.  We then use the ages to discuss possible connections between the distinct subclusters, including a possible explanation similar to the sequential star formation in Sco-Cen OB Association, as inferred by \citet{Krause2018}.  A deeper understanding of the star formation history of the Serpens Molecular Cloud and any causal connections between different regions will require spectroscopy, especially of infrared-selected members that are not optically visible, and analysis of star and gas motions.

\acknowledgements

We thank the anonymous referee for the helpful comments and
suggestions that helped to significantly improve the analysis and the clarity of the paper. We thank Marina Kounkel for discussion of APOGEE temperature and ages of Serpens and Kevin Covey for discussion of temperature scales.

We thank Jennifer Hatchell, Carlo Manara, Doug Johnstone, Michael Dunham, Anupam Bhardwaj, Jessy Jose, and Zhen Yuan for their effort in our initial paper that developed the sample used here. 

XYZ and GJH are supported by general grants 12173003 and 11773002 awarded by the Natural Science Foundation of China. YL acknowledges financial supports by the Natural Science Foundation of China (Grant No. 11973090) and by the Natural Science Foundation of Jiangsu Province of China (Grant No. BK20181513).

This work has made use of data from the European Space Agency (ESA) mission
{\it Gaia} (\url{https://www.cosmos.esa.int/gaia}), processed by the {\it Gaia}
Data Processing and Analysis Consortium (DPAC, \url{https://www.cosmos.esa.int/web/gaia/dpac/consortium}). Funding for the DPAC
has been provided by national institutions, in particular the institutions
participating in the {\it Gaia} Multilateral Agreement.
This publication makes use of data products from the Two Micron All Sky Survey, which is a joint project of the University of Massachusetts and the Infrared Processing and Analysis Center/California Institute of Technology, funded by the National Aeronautics and Space Administration and the National Science Foundation.
This publication makes use of data products from the Wide-field Infrared Survey Explorer, which is a joint project of the University of California, Los Angeles, and the Jet Propulsion Laboratory/California Institute of Technology, funded by the National Aeronautics and Space Administration.
We acknowledge with thanks the variable star observations from the AAVSO International Database contributed by observers worldwide and used in this research.  This research has made use of NASA's Astrophysics Data System.
This research has made use of the SIMBAD database, operated at CDS, Strasbourg, France.
This research has made use of the Spanish Virtual Observatory (\url{http://svo.cab.inta-csic.es}) supported from the Spanish MICINN/FEDER through grant AyA2017-84089.  This research made use of Astropy, a community-developed core Python package for Astronomy \citep{Astropy2013}.

\pagebreak

\appendix
\restartappendixnumbering
\twocolumngrid

\section{Color-color relations}
\label{appendix:sec:other_color_relations}

\begin{figure*}[ht]
    \centering
    \includegraphics[width=\textwidth]{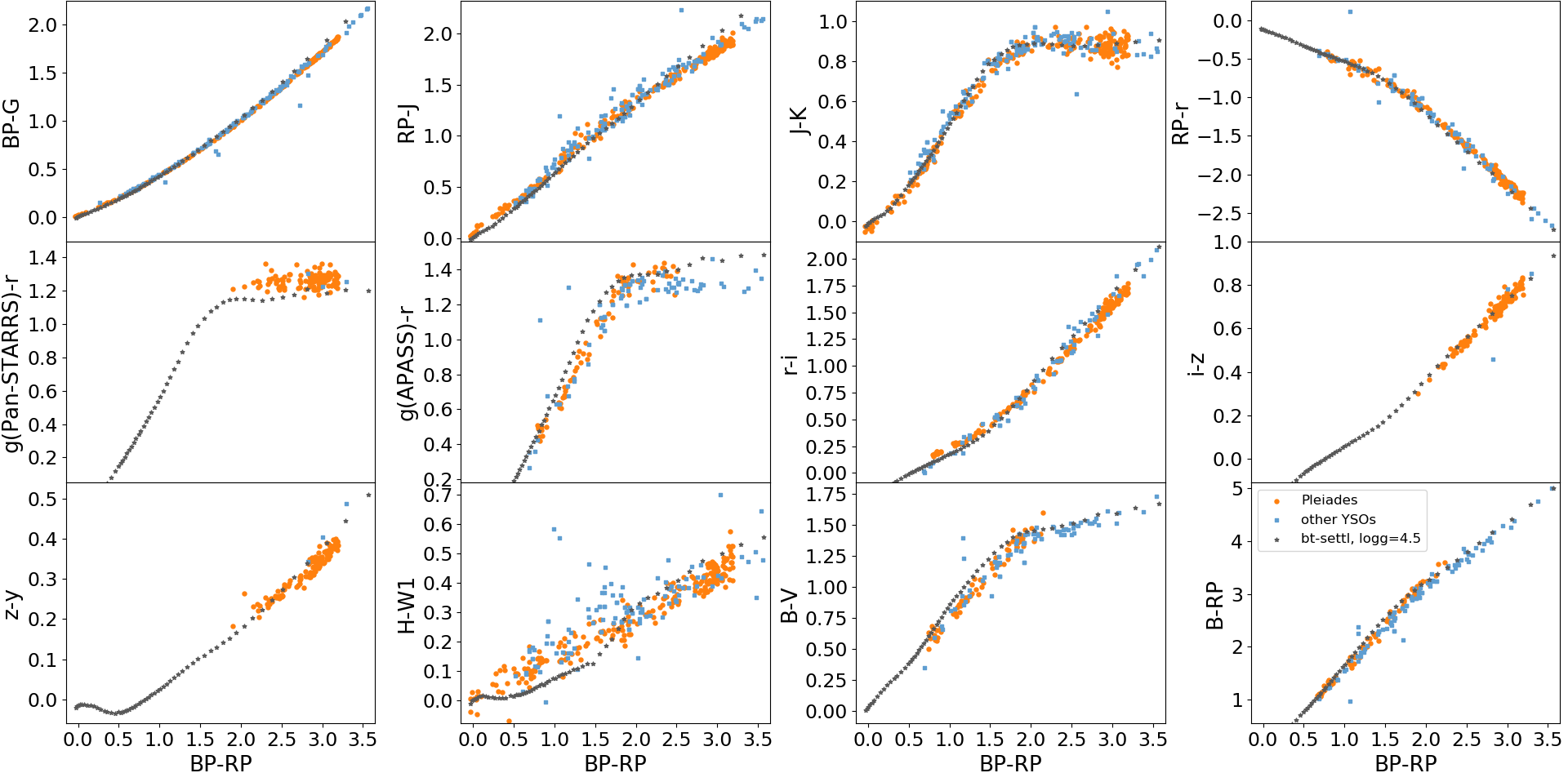}
    \caption{Relations between various colors and $BP-RP$.  Red, cyan and blue dots represent colors from Pleiades stars, colors from other YSOs (\citet{Pecaut2013}, \citet{Gagne2018}, \citet{Manara2013}, \citet{Manara2017}), and colors calculated through BT-Settl spectra. A few stars in \citet{Pecaut2013} may contain a dusty disk, according to $H-W1$.}
    \label{fig:more_color_color}
\end{figure*}

In Section \ref{sec:data}, we describe empirical color-color relationships and compare them to models. Here we provide supplemental figures and a Table presenting the colors used in our fits.

\startlongtable
\begin{deluxetable*}{ccccccccccccccc}
\label{table:color_color_table}
\tablecaption{Color-color table}
\tablehead{\colhead{T/K} & \colhead{BP-RP} & \colhead{J-RP} & \colhead{H-RP} & \colhead{K-RP} & \colhead{W1-RP} & \colhead{W2-RP} & \colhead{g-RP} & \colhead{r-RP} & \colhead{i-RP} & \colhead{z-RP} & \colhead{y-RP} & \colhead{B-RP} & \colhead{V-RP} & \colhead{g'-RP}}
\startdata
3000 & 3.17  & -1.93 & -2.54 & -2.83 & -3.02 & -3.23 & 3.51 & 2.32 & 0.55 & -0.21 & -0.61 & -    & -    & -     \\
3100 & 3.04  & -1.86 & -2.47 & -2.74 & -2.92 & -3.12 & 3.38 & 2.17 & 0.53 & -0.2  & -0.56 & 4.24 & 2.76 & 3.39  \\
3200 & 2.90   & -1.79 & -2.40  & -2.67 & -2.83 & -3.02 & 3.25 & 2.03 & 0.50  & -0.17 & -0.52 & 4.17 & 2.64 & 3.33  \\
3300 & 2.79  & -1.73 & -2.36 & -2.62 & -2.76 & -2.94 & 3.15 & 1.91 & 0.49 & -0.15 & -0.48 & 4.09 & 2.54 & 3.26  \\
3400 & 2.70   & -1.69 & -2.32 & -2.57 & -2.71 & -2.87 & 3.05 & 1.81 & 0.47 & -0.13 & -0.44 & 4.01 & 2.45 & 3.19  \\
3500 & 2.60   & -1.64 & -2.29 & -2.53 & -2.66 & -2.80  & 2.96 & 1.71 & 0.46 & -0.11 & -0.40  & 3.93 & 2.36 & 3.12  \\
3600 & 2.50   & -1.59 & -2.25 & -2.49 & -2.61 & -2.73 & 2.86 & 1.62 & 0.45 & -0.09 & -0.37 & 3.83 & 2.26 & 3.03  \\
3700 & 2.40   & -1.54 & -2.21 & -2.44 & -2.55 & -2.66 & 2.76 & 1.52 & 0.43 & -0.06 & -0.33 & 3.71 & 2.16 & 2.93  \\
3800 & 2.29  & -1.49 & -2.17 & -2.39 & -2.49 & -2.58 & 2.65 & 1.43 & 0.42 & -0.03 & -0.28 & 3.58 & 2.06 & 2.82  \\
3900 & 2.17  & -1.43 & -2.12 & -2.33 & -2.42 & -2.49 & 2.53 & 1.33 & 0.41 & 0.00     & -0.23 & 3.44 & 1.94 & 2.69  \\
4000 & 2.05  & -1.37 & -2.06 & -2.25 & -2.34 & -2.39 & 2.41 & 1.23 & 0.40  & 0.02  & -0.18 & 3.27 & 1.82 & 2.55  \\
4100 & 1.93  & -1.30  & -1.98 & -2.17 & -2.25 & -2.28 & 2.28 & 1.13 & 0.39 & 0.05  & -0.14 & 3.10  & 1.70  & 2.41  \\
4200 & 1.82  & -1.24 & -1.91 & -2.08 & -2.16 & -2.17 & 2.15 & 1.05 & 0.38 & 0.09  & -0.09 & 2.93 & 1.59 & 2.26  \\
4300 & 1.72  & -1.17 & -1.83 & -1.99 & -2.07 & -2.07 & 2.03 & 0.97 & 0.37 & 0.11  & -0.05 & 2.77 & 1.49 & 2.13  \\
4400 & 1.63  & -1.12 & -1.75 & -1.91 & -1.98 & -1.97 & 1.92 & 0.91 & 0.36 & 0.14  & -0.01 & 2.63 & 1.40  & 2.00     \\
4500 & 1.55  & -1.06 & -1.68 & -1.82 & -1.89 & -1.88 & 1.81 & 0.85 & 0.36 & 0.16  & 0.02  & 2.49 & 1.31 & 1.89  \\
4600 & 1.47  & -1.02 & -1.60  & -1.75 & -1.81 & -1.79 & 1.71 & 0.80  & 0.35 & 0.18  & 0.05  & 2.37 & 1.24 & 1.79  \\
4700 & 1.40   & -0.97 & -1.54 & -1.67 & -1.73 & -1.71 & 1.62 & 0.76 & 0.35 & 0.20   & 0.07  & 2.25 & 1.17 & 1.70   \\
4800 & 1.34  & -0.93 & -1.47 & -1.60  & -1.66 & -1.63 & 1.54 & 0.72 & 0.35 & 0.22  & 0.10   & 2.15 & 1.11 & 1.61  \\
4900 & 1.28  & -0.89 & -1.41 & -1.53 & -1.59 & -1.56 & 1.46 & 0.69 & 0.34 & 0.23  & 0.12  & 2.05 & 1.06 & 1.53  \\
5000 & 1.23  & -0.85 & -1.34 & -1.46 & -1.52 & -1.49 & 1.38 & 0.66 & 0.34 & 0.24  & 0.14  & 1.96 & 1.01 & 1.45  \\
5200 & 1.13  & -0.78 & -1.23 & -1.33 & -1.39 & -1.36 & 1.23 & 0.60  & 0.33 & 0.26  & 0.18  & 1.79 & 0.91 & 1.32  \\
5400 & 1.04  & -0.72 & -1.12 & -1.22 & -1.27 & -1.24 & 1.10  & 0.56 & 0.33 & 0.28  & 0.21  & 1.63 & 0.83 & 1.20   \\
5600 & 0.95  & -0.66 & -1.02 & -1.11 & -1.16 & -1.13 & 0.97 & 0.52 & 0.33 & 0.30   & 0.24  & 1.49 & 0.75 & 1.09  \\
5800 & 0.88  & -0.60  & -0.92 & -1.00    & -1.05 & -1.02 & 0.84 & 0.48 & 0.32 & 0.31  & 0.27  & 1.36 & 0.68 & 0.99  \\
6000 & 0.80   & -0.55 & -0.83 & -0.90  & -0.95 & -0.92 & 0.72 & 0.45 & 0.32 & 0.32  & 0.29  & 1.24 & 0.62 & 0.89  \\
6200 & 0.74  & -0.50  & -0.74 & -0.81 & -0.86 & -0.83 & -    & 0.42 & 0.32 & -     & -     & 1.12 & 0.56 & 0.81  \\
6400 & 0.67  & -0.46 & -0.66 & -0.73 & -0.77 & -0.75 & -    & 0.39 & 0.31 & -     & -     & 1.01 & 0.51 & 0.73  \\
6600 & 0.61  & -0.41 & -0.59 & -0.64 & -0.68 & -0.66 & -    & 0.37 & 0.31 & -     & -     & 0.91 & 0.46 & 0.65  \\
6800 & 0.55  & -0.37 & -0.51 & -0.57 & -0.60  & -0.59 & -    & 0.34 & 0.31 & -     & -     & 0.81 & 0.41 & 0.58  \\
7000 & 0.49  & -0.33 & -0.45 & -0.50  & -0.53 & -0.52 & -    & 0.32 & 0.30  & -     & -     & 0.72 & 0.36 & 0.52  \\
7500 & 0.36  & -0.25 & -0.31 & -0.34 & -0.36 & -0.36 & -    & 0.28 & 0.30  & -     & -     & 0.52 & 0.26 & 0.38  \\
8000 & 0.24  & -0.19 & -0.20  & -0.23 & -0.24 & -0.24 & -    & 0.25 & 0.29 & -     & -     & 0.35 & 0.18 & 0.26  \\
8500 & 0.14  & -0.13 & -0.12 & -0.14 & -0.14 & -0.15 & -    & 0.22 & 0.29 & -     & -     & 0.21 & 0.12 & 0.18  \\
9000 & 0.06  & -0.10  & -0.07 & -0.08 & -0.08 & -0.09 & -    & 0.21 & 0.28 & -     & -     & 0.10  & 0.06 & 0.11  \\
9500 & 0     & -0.07 & -0.03 & -0.04 & -0.03 & -0.04 & -    & -    & -    & -     & -     & -    & -    & -     \\
\enddata
\end{deluxetable*}

\section{Fitting additional young stars}
\label{appendix:fit_additional_YSOs}

As mentioned in Section \ref{subsec:selection_Pleiades}, some additional young stars are used to expand temperature range of Pan-STARRS photometry. These stars include: a) $< \sim30$ Myr $\beta$ Pic, $\epsilon$ Cha, $\eta$ Cha, and TW Hya from BANYAN XI and XIII \citep{Gagne2018, Gagne2018b}; b) $\sim1000$ stars selected from \citet{KounkelCovey2019}. The \citet{KounkelCovey2019} catalog contains $\sim0.3$ million stars within 1 kpc and with ages from $\sim1$ Myr to $\sim10$ Gyr. To select stars younger than or similar to Pleiades, we first pick stars younger than 300 Myr and within 1000 pc \footnote{The distance cut is based on \textit{Gaia} EDR3 which may be different from \textit{Gaia} DR2 distance used in \citet{KounkelCovey2019}.} from \citet{KounkelCovey2019}. Then, only stars with photometry of all pass-bands in Section \ref{sec:data} are selected, leaving $\sim12000$ stars. Finally, since expanding temperature range of Pan-STARRS photometry does not need a large amount of stars (e.g., the Pleiades sample in this method only includes $\sim300$ stars), we further select $\sim1000$ stars from the $\sim12000$ stars, which is done by picking 1000 uniformly distributed numbers between 0 and 3.2 (the $BP-RP$ range of this fitting method), and finding stars with $BP-RP$ locating at these numbers. Such selection assures that the collected stars are basically uniformly distributed in the color range of the fitting, except the bluest end which is too bright for Pan-STARRS photometry.

\begin{figure}[!t]
    \centering
    \includegraphics[width=0.9\columnwidth]{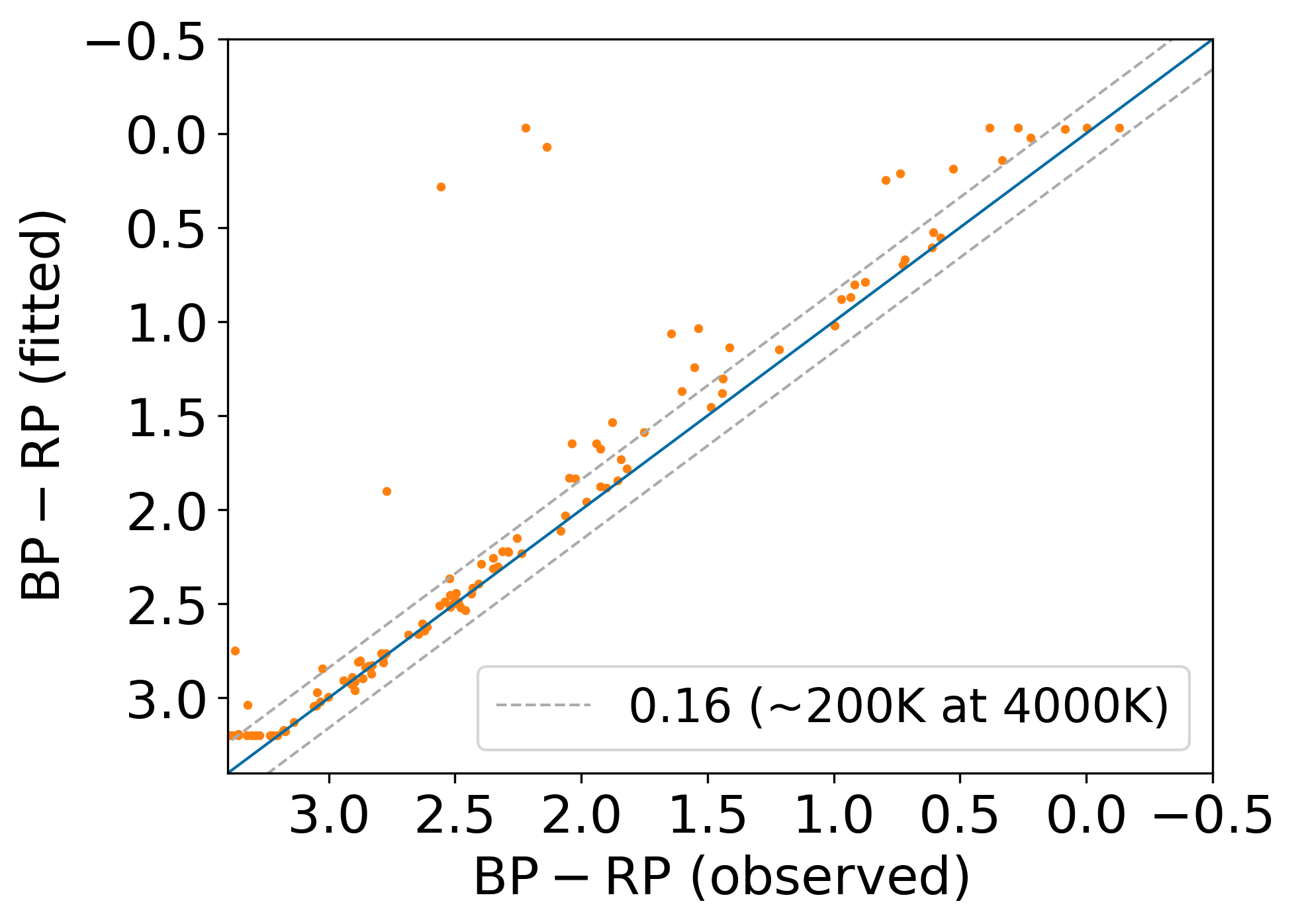}
    \caption{Comparison between observed $BP-RP$ (x-axis) and fitted $BP-RP$ (y-axis) of BANYAN YSOs.}
    \label{fig:use_Pleiades_fit_BANYAN}
\end{figure}

\begin{figure}[!t]
    \centering
    \includegraphics[width=0.9\columnwidth]{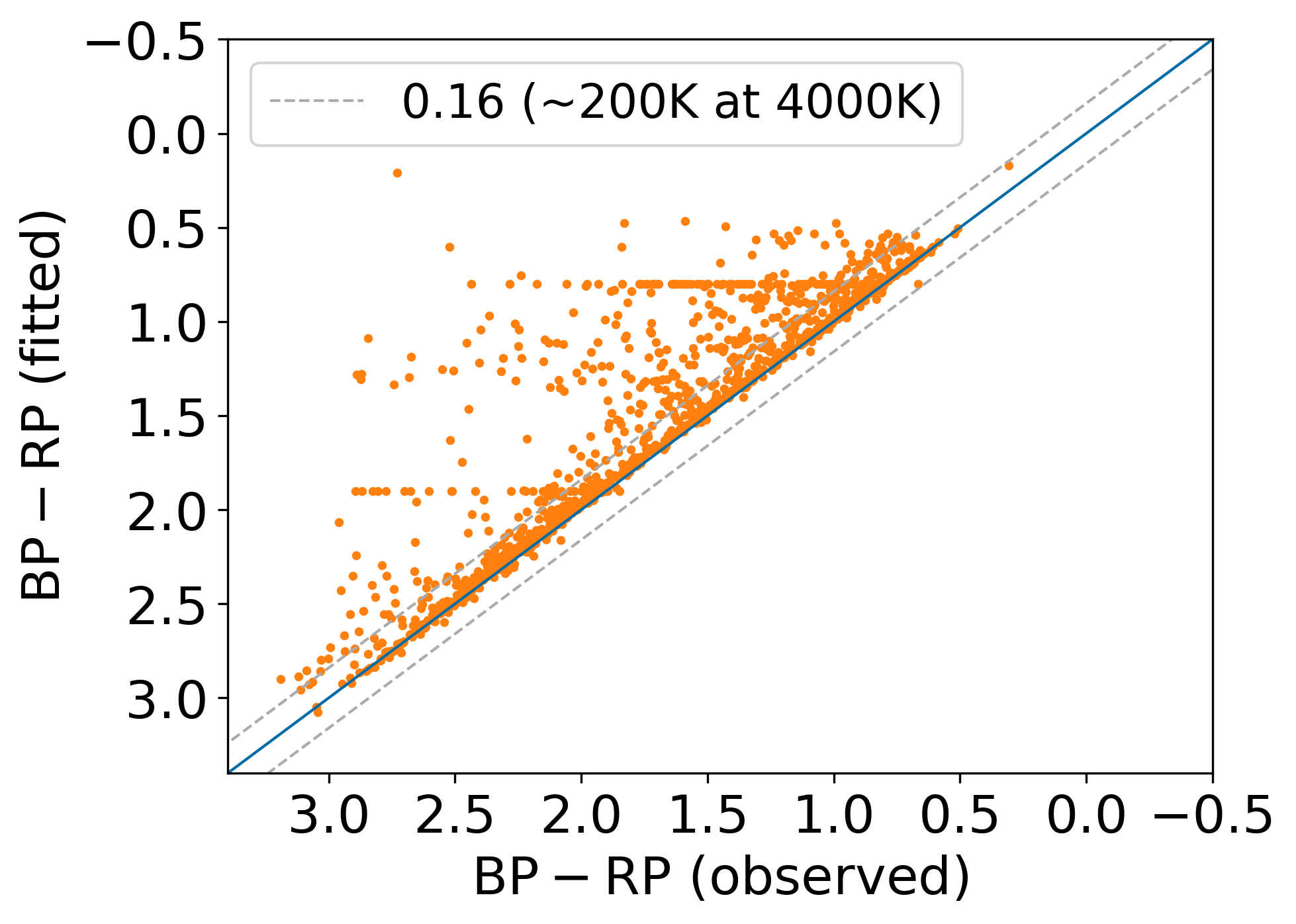}
    \caption{Comparison between observed $BP-RP$ (x-axis) and fitted $BP-RP$ (y-axis) of \citet{Kounkel2019} young stars. Extinction in the fitting is restricted to $A_V \geq 0$.}
    \label{fig:use_Pleiades_fit_K19}
\end{figure}

To investigate whether Pleiades fits the additional young stars well, we use Pleiades to fit these stars, following the method in Section \ref{sec:fitting_method}. Figure \ref{fig:use_Pleiades_fit_BANYAN} and \ref{fig:use_Pleiades_fit_K19} show the fitted result of BANYAN and \citet{KounkelCovey2019} stars.
Because there is no APOGEE catalog containing most of these additional stars \footnote{\citet{KounkelCovey2019} includes stars in Orion and IC 348 where APOGEE has observed, while most stars in \citet{KounkelCovey2019} are not observed by APOGEE.}, the color $BP-RP$ is used for comparison instead of $T_{\rm eff}$.

As shown in Figure \ref{fig:use_Pleiades_fit_BANYAN}, Pleiades fits well to the cooler YSOs ($BP-RP>2$, corresponding to $\sim4000$ K). Among the three stars with the bluest fitted $BP-RP$, two have a protoplanetary disk and one has a nearby ($\sim6$ arcsec) bright source that limits the accuracy of \textit{Gaia} and 2MASS photometry and may contaminate 2MASS photometry. Pleiades also fits well to the hotter YSOs ($BP-RP<1$, corresponding to $\sim5500$ K). Three stars at $BP-RP \sim 0.7$ show flux excess in $W1$ and $W2$, thus have bluer fitted $BP-RP$. Their fitted $BP-RP$ is close to observed values after removing WISE data. Stars at $1<BP-RP<2$ tend to have bluer fitted $BP-RP$, which is likely due to surface gravity difference.

Figure \ref{fig:use_Pleiades_fit_K19} also shows a good consistency between fitted $BP-RP$ and observed $BP-RP$. Some stars may have obvious extinction and/or disk, thus have bluer fitted $BP-RP$. The line structure at $(BP-RP)_{\rm fit} \sim 1.9$ and $(BP-RP)_{\rm fit} \sim 0.8$ corresponds to $BP-RP$ range of Pleiades Pan-STARRS and APASS data. The fitted $BP-RP$ is hardly redder than the observed $BP-RP$, due to the $A_V$ constraint ($>=0$) in the fitting.
We then select additional young stars based on the result above, as described in Section \ref{subsec:selection_Pleiades}.

\section{Age estimate for Collinder 359}
\label{appendix:coll359}

\begin{figure}[!t]
    \centering
    \includegraphics[width=\columnwidth]{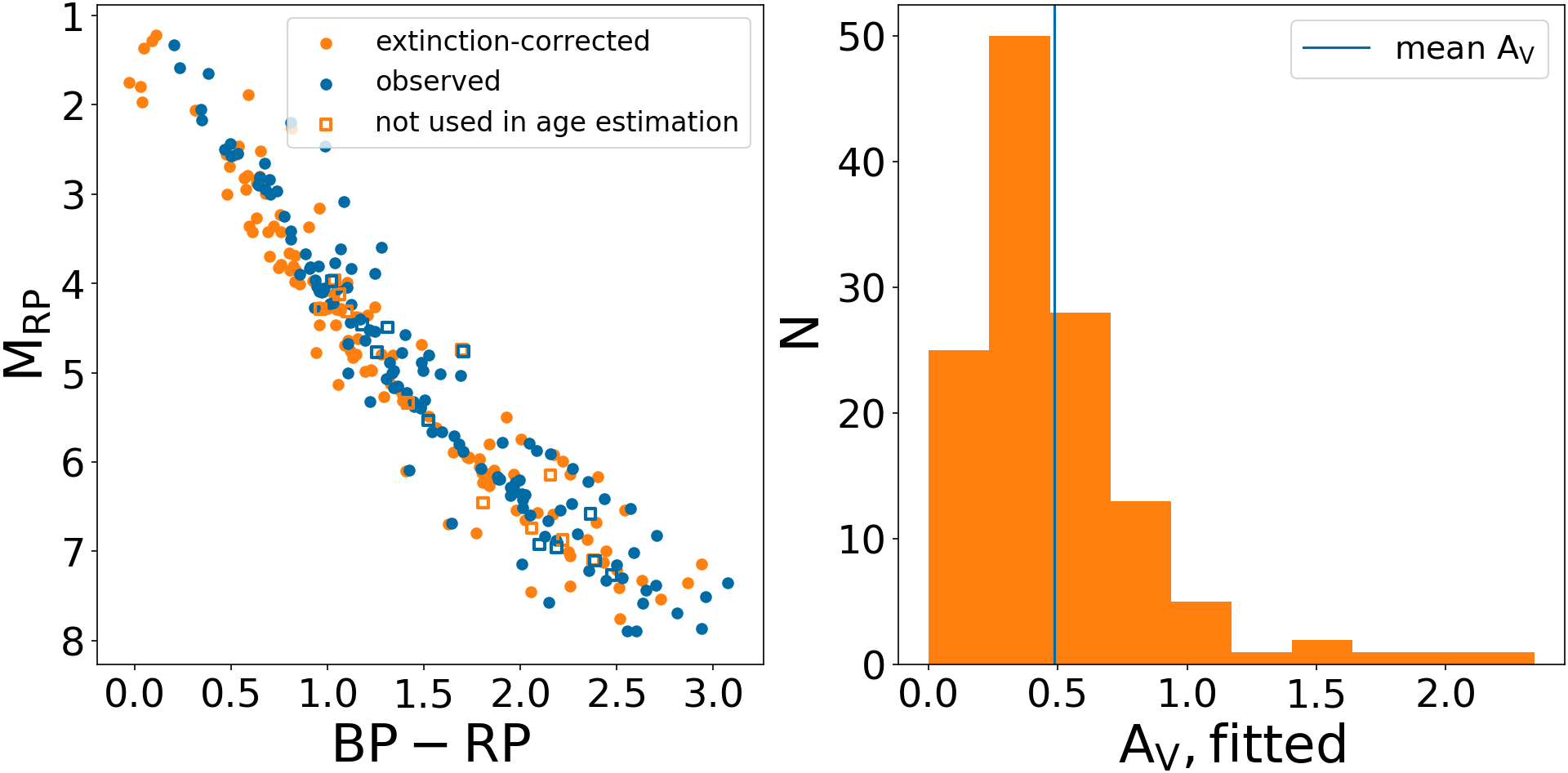}
    \caption{{\textit{Gaia} EDR3} color-magnitude diagram (left panel) and distribution of best-fit $A_V$  (right panel) of Collinder 359. The color-magnitude diagram includes the observed (blue) and extinction-corrected (orange) colors and magnitudes. Unfilled symbols are stars which are not used in calculating ages, and these stars are not included in the right panel.}
    \label{fig:CMD_Col359}
\end{figure}

Collinder 359 is a young open cluster located $\sim130$ pc away from Serpens Main. The distance to Collinder 359 is $\sim560$ pc \citep{Cantat-Gaudin2018}. In Section \ref{subsec:subgroup_dynamic}, we evaluate the possible connection between Collinder 359 and the ongoing star formation in Serpens, since they are physically located 130 pc away from each other.  Collinder 359 has been previously estimated to have an age of $\sim60$ Myr \citep{Lodieu2006} and $\sim28$ Myr \citep{Kharchenko2013}.

We re-evaluate the age of Collinder 359 as {21 Myr} from the \citet{Feiden2016} non-magnetic tracks, in a method consistent with the approach applied to the Serpens sample of young stars. The age is estimated from {127 stars selected from 138 members in \citet{Cantat-Gaudin2018} with membership probability $P_{\rm memb} \geq 0.9$, with 11 stars excluded due to their visual binaries that are unresolved by 2MASS.} Figure \ref{fig:CMD_Col359} shows color-magnitude diagram and $A_V$ distribution of these Collinder 359 stars. After fitting a three-degree polynomial to the color-magnitude diagram, the rms is {0.41} for the observed CMD, {0.40} for the extinction-corrected CMD based on SED fitting result, and {0.38} for the observed CMD when using a single distance for all stars. The SED fitting result does not significantly affect the rms in color-magnitude diagram, due to the tight $A_V$ distribution of Collinder 359. The average fitted $A_V\sim$ {0.5 mag} is consistent with $A_V\sim0.5$ mag estimated by \citet{Kharchenko2013}.

\clearpage
\pagebreak

\bibliography{export-bibtex}

\end{document}